\documentclass[12pt,a4paper]{article}
\usepackage{amsmath}
\usepackage{multirow,booktabs}
\usepackage{graphicx}
\usepackage{amssymb}
\usepackage{amsthm}
\usepackage{enumitem}
\usepackage{mathtools}
\usepackage{setspace} 
\usepackage{tikz}
\usepackage{caption}
\usepackage[titletoc,title]{appendix}

\usepackage{soul}

\title{{\sc Analysis of Left Truncated and Right Censored Competing Risks Data}}
\author{Debasis Kundu$^1$ \& Debanjan Mitra$^2$  \& Ayon Ganguly$^3$}
\date{}

\begin{document}

\maketitle
\parskip = 1em

\begin{abstract}
In this article, the analysis of left truncated and right censored competing risks data is carried out, under the assumption of
the latent failure times model.  It is assumed that there are two competing causes of failures, although most of the results can
be extended for more than two causes of failures. The lifetimes corresponding to the competing causes of failures are assumed
to follow Weibull distributions with the same shape parameter but different scale parameters.  The maximum likelihood estimation
procedure of the model parameters is discussed, and confidence intervals are provided using the bootstrap approach.  When the
common shape parameter is known, the maximum likelihood estimators of the scale parameters can be obtained in explicit forms, and
when it is unknown we provide a simple iterative procedure to compute the maximum likelihood estimator of the shape parameter.
The  Bayes estimates and the associated credible intervals of unknown parameters are also addressed under a  very flexible set of
priors on the shape and scale parameters.   Extensive Monte Carlo simulations are performed to compare the performances of the
different methods.  A numerical example is provided for illustrative purposes.  Finally the results have been extended when the
two competing causes of failures are assumed to be independent Weibull distributions with different shape parameters.
\end{abstract}

\vspace{0.5 in}



\noindent $^1$ Department of Mathematics and Statistics, Indian Institute of Technology
Kanpur, Uttar Pradesh 208016, India.  Corresponding author.  E-mail:kundu@iitk.ac.in

\noindent $^2$ Operations Management, Quantitative Methods and
Information Systems Area, Indian Institute of Management Udaipur, India.

\noindent$^3$ Department of Mathematics, Indian Institute of Technology Guwahati, Guwahati, Assam 781039, India.

\doublespacing

\section{\sc Introduction}

In the analysis of reliability data or in medical studies, the failure of an item or an individual may be attributable to more
than one cause or factor.  These `risk factors' in some sense compete with  each other for the failure of the experimental unit.
An investigator is often interested in the assessment of a specific risk in the presence of other risk factors.  In the
statistical literature it is well known as the competing risks model.  In analyzing the competing risks model, it is assumed that
the data consists of a failure time and an indicator denoting the cause of failure.  An extensive amount of work has been carried
out on the analysis of competing risks data both under the parametric and non-parametric set-up.  See for example Crowder (2001),
David and Moeschberger (1978) and the references cited therein for different issues related to the competing risks problems.

The analysis of lifetime data in the competing risks framework can be performed in two different ways: one can either adapt the
latent failure times model approach as suggested by Cox (1959), or use the cause specific hazard function model as suggested by
Prentice et al. (1978).  In the non-parametric set up no specific lifetime distribution is assumed.  For the parametric set up it
is assumed that different causes follow some specific parametric distribution, namely exponential, gamma, Weibull etc.  It is
observed by Kundu (2004) that when the assumed model for the lifetime is either exponential or Weibull, the two above approaches
lead to the same likelihood function, hence provide the same set of estimators of the unknown parameters.  Although the
interpretations of the model parameters are quite different.

The problem addressed in this paper was mainly motivated from a real life example mentioned in Hong, Meeker and McCalley (2009),
and it can be stated as follows.  There are approximately 150,000 high-voltage power transformers which were installed at
different time points in the past and they are in service in different parts of U.S.  The energy company started record keeping
only in 1980.  The complete information on transformers installed after 1980 are available. Moreover, the complete information on
transformers which were installed before 1980 but failed after 1980 are also available.  However, no information is available on
those units which were installed before 1980 and failed before 1980.  The authors had the access to the data till 2008.
Therefore, all the units which have not failed till 2008 are right censored.  The data of this type are known as the left
truncated right censored data.  Due to confidentiality reason the authors did not provide the exact data, but they provided the
classical analysis of the data set based on the assumption that the lifetime distribution of the transformers follow a
two-parameter Weibull distribution.  

Recently, Balakrishnan and Mitra (2012) mimicked the lifetime of the transformer data of Hong, Meeker and McCalley (2009) and
provided a detailed analysis of the model.  In this connection see also Balakrishnan and Mitra (2011, 2014), where the authors
considered different other lifetime distributions and proposed to use the expectation maximization (EM) algorithm to compute the
maximum likelihood estimators (MLEs) of the unknown parameters and also provided the confidence intervals of the unknown
parameters based on missing information principle.  Very recently Kundu and Mitra (2016) considered the same problem from the
Bayesian perspective. In this paper we consider the same problem as mentioned in Hong, Meeker and McCalley (2009) with the further
assumption that each transformer can fail due to some cause, for example (i) excessive load or (ii) excessive heating etc.  If the
lifetime of an unit is available then the corresponding cause of failure is also known.  We call this type of data as the left
truncated right censored competing risks data. For notational simplicity it is assumed that we have only two causes of failures
although all the results provided here can be easily generalized for any number of causes.

As in Balakrishnan and Mitra (2012) we have mimicked the lifetime of the transformer data of Hong, Meeker and McCalley (2009) with
a possible causes of failure. The data set is presented in the Appendix A.  Here $\nu$ = 1 indicates that the transformer was
installed after 1980, and $\nu$ = 0 indicates that it was installed before 1980 and it did not fail till 1980.  Further, $\delta$
= 1 or 2 indicates that the transformer has failed due to Cause 1 or Cause 2, respectively, and $\delta$ = 0 implies it did not
fail till 2008.  The main aim of this paper is to provide the detailed analysis of this left truncated right censored competing
risks data set.

To analyze this data set it is assumed that the competing causes of failures follow Cox's latent failure time model assumptions.
Moreover, it is further assumed that the failure time distributions of both the causes follow two-parameter Weibull distribution
with the common shape parameter but different scale parameters.  First we obtain the MLEs of the unknown parameters.  It is
observed that when the common shape parameter is known the MLEs of the scale parameters can be obtained in explicit forms.  When
the common shape parameter is unknown, first we obtain the MLE of the shape parameter by solving a simple non-linear equation, and
then we obtain the MLEs of the scale parameters in explicit forms.  We have proposed to use the parametric bootstrap method for
constructing the confidence intervals of the unknown parameters.  

We further provide the Bayesian analysis of the unknown parameters.  When the common shape parameter is known we have assumed a
very flexible conjugate Dirichlet-Gamma (DG) prior on the scale parameters.  In this case the Bayes estimates and the associated
credible interval can be obtained in explicit form.  When the common shape parameter is unknown, no specific form of prior on the
shape parameter is assumed.  It is assumed that the shape parameter has a prior which has a log-concave density function.  In this
case the Bayes estimates cannot be obtained in explicit forms.  We propose to use the importance sampling procedure to compute the
Bayes estimates and also to construct the associated credible intervals.  Extensive simulations have been performed to compare the
performances of the different methods and one data analysis has been performed for illustrative purposes.  Finally we extend the
results when the shape parameters of the two competing causes of failures are not assumed to be the same.  We provide the
classical and Bayesian inference under this generalized assumption and reanalyze the same data set for illustrative purposes.  

It may be mentioned that although quite a bit of work has been done so far on the analysis of left truncated right censored data,
nobody has provided the analysis in presence of competing risks.  In particular, Balakrishnan and Mitra (2012) and Kundu and 
Mitra (2016) provided the classical and Bayesian analysis, respectively,  of the left truncated right censored data when 
the lifetime distribution of the experimental units follow Weibull distribution without any presence of competing risks.
In this paper we provide both the classical and Bayesian
inference for the left truncated right censored competing risks data under a fairly general set of priors, and that is the major
contribution of this paper.  Although, we have assumed that the competing causes of failures follow Weibull distributions, similar
procedures may be developed for other distributions also.

The rest of the paper is organized as follows. We describe the basic model and the notations used in the paper in Section 2. In
Section 3, we discuss the maximum likelihood estimation procedure of the model parameters  and also the construction of the
associated confidence intervals based on parametric bootstrap approach.  Next, we discuss the Bayesian analysis of this problem in
Section 4, where we provide the Bayes estimates and the associated credible intervals.  In Section 5, we present the Monte Carlo
simulation results to compare the performances of the different methods proposed here, and the analysis of one data set is
provided in Section 6.  In Section 7, we provide the classical and Bayesian inference of the unknown parameters when the shape
parameters need not be equal, and finally, we conclude the paper with some remarks in Section 8.

\section{\sc Model description and notation}

Experimental units are put on a life test at different time points.  Let the lifetime of an experimental unit be denoted
by a random variable $T$.  For each experimental unit there is one left truncation time point say $\tau_L$, which may depend on the
experimental unit.  Suppose an experimental unit has been put on a life test at the time point 0, and it has the left truncation
time point $\tau_L$.  If $\tau_L > 0$, then the information about the failure time $T$ of the experimental unit is available 
if $T > \tau_L$, 
otherwise no information is available about $T$.  On the other hand if $\tau_L < 0$, the information about $T$ is always available.
If an item has been put on a life test before $\tau_L$, and it is failed after
$\tau_L$, then the failure time is known as the truncated failure time.  
If an experimental unit has been put on a test before the left
truncation point $\tau_L$ or it has been put on a test after $\tau_L$, it may be censored at the right censoring point $\tau_R >
\tau_L$.  The right censoring point $\tau_R$ may also depend on the experimental unit.  Therefore, if an experimental unit has been
put on a test at the time point 0, and $\tau_L > 0$, then the exact failure time is known if $\tau_L < T < \tau_R$.  Similarly, for
$\tau_L < 0$, the exact failure time is known provided $T < \tau_R$.   If the exact failure time of an experimental unit is
observable, then the corresponding cause of failure is also known. 
For
example, in case of transformer-example as provided in the previous section, $\tau_L$ for a particular transformer is 1980 minus
the year of installment of the transformer, and $\tau_R$ is 2008 minus the  year of installment of the transformer.  The necessary information of an experimental unit is
available only if it fails after $\tau_L$, or it is being censored after $\tau_L$.  Therefore, the information regarding the
number of failures before the left truncation point is not available.  We use the following notations for the rest of the paper.

\noindent $T_{ji}$: latent failure time of the $i$th unit under cause $j$, $i$=1,2,...,$n$, $j$=1,2. \newline
\noindent $\tau_{iL}$: left truncation time for the $i$-th unit.  \newline
\noindent $\tau_{iR}$: right censoring time for the $i$-th unit.  \newline
\noindent $T_i$: lifetime of the $i$-th unit.   \newline
\noindent $I_j$: set of indices of failures due to cause $j$, $j$=1,2. \newline
\noindent $I_0$: set of indices of censored observations. \newline
\noindent $|I_j|$: cardinality of $I_j$. We assume that $|I_j| = m_j$, $j$=1,2 and $m = m_1+m_2$. \newline
\noindent $\delta_i$: indicator variable for the $i$th unit (1 if it fails from cause 1; 2 if it fails from
cause 2; 0 if it is censored). \newline
\noindent $\nu_i$: truncation indicator. It is 1 if $i$th unit is not truncated; 0 if it is truncated.
\newline
\noindent Weibull$(\alpha, \lambda)$: Weibull random variable with probability density function $\alpha
\lambda x^{\alpha - 1}e^{-\lambda x^{\alpha}}$; $x>0$. 

\noindent It is assumed that $(T_{1i}, T_{2i})$, for $i = 1, \ldots, n$, are $n$ independent identically distributed random 
vectors.  $T_{1i}$ and $T_{2i}$ are independent for all $i = 1, \ldots, n$, and $T_i = \min\{T_{1i}, T_{2i}\}$, see Cox (1959).
It is further assumed here that $T_{1i}$ follows ($\sim$) Weibull $(\alpha, \lambda_1)$ and $T_{2i} \sim$ 
Weibull$(\alpha, \lambda_2)$ distribution.

\section{\sc Likelihood inference}

\subsection{\sc Maximum Likelihood Estimators}

It is assumed that all the units are put on a test at the time point 0, otherwise, necessary adjustment needs to be made.
For the observation $\{(t_i, \delta_i, \nu_i); i = 1, \ldots, n\}$, the likelihood contribution of an experimental 
unit for different values of $\delta$ and $\nu$ are as follows: 

\noindent Case 1: $\displaystyle \alpha \lambda_1 t_i^{\alpha-1} e^{-(\lambda_1+\lambda_2)t_i^{\alpha}}$, when $\delta_i$ = 1, $\nu_i$ = 1 
\vskip 0.05in
\noindent Case 2: $\alpha \lambda_2 t_i^{\alpha-1} e^{-(\lambda_1+\lambda_2)t_i^{\alpha}}$, when $\delta_i$ = 2, $\nu_i$ = 1 \vskip 0.05in
\noindent Case 3: $e^{-(\lambda_1+\lambda_2)t_i^{\alpha}}$, when $\delta_i$ = 0, $\nu_i$ = 1 
\vskip 0.05in
\noindent Case 4: $\displaystyle \frac{\alpha \lambda_1 t_i^{\alpha-1} e^{-(\lambda_1+\lambda_2)t_i^{\alpha}}}{e^{-(\lambda_1+\lambda_2){\tau_{iL}}^{\alpha}}}$, when $\delta_i$ = 1, $\nu_i$ = 0 
\vskip 0.05in
\noindent Case 5: $\displaystyle \frac{\alpha \lambda_2 t_i^{\alpha-1} e^{-(\lambda_1+\lambda_2)t_i^{\alpha}}}{e^{-(\lambda_1+\lambda_2){\tau_{iL}}^{\alpha}}}$, when $\delta_i$ = 2, $\nu_i$ = 0 
\vskip 0.05in
\noindent Case 6: $\displaystyle \frac{e^{-(\lambda_1+\lambda_2)t_i^{\alpha}}}{e^{-(\lambda_1+\lambda_2){\tau_{iL}}^{\alpha}}}$, when $\delta_i$ = 0, $\nu_i$ = 0. 

We will explain Case 1 and Case 4 in details.  Rest will follow along the same manner.  Case 1: In this case since $\nu_i$ = 1, it means
the unit has not been left truncated, and since $\delta_i$ = 1, it implies $T_{1i} = t_i$, and $T_{2i} > t_i$.  Therefore, the likelihood
contribution becomes $\displaystyle P(t_i < T_{1i} < t_i+dt_i, T_{2i} > t_i) = 
\alpha \lambda_1 t_i^{\alpha-1} e^{-\lambda_1 t_i^{\alpha}} e^{-\lambda_2 t^{\alpha}} dt_i$.
Similarly, for Case 4, 
since $\nu_i$ = 0, it means that the unit has been left truncated.  Hence, we know that $T_i = \min\{T_{1i}, T_{2i}\} > \tau_{iL}$.  Moreover,
$\delta_i$ = 1, implies that $T_{1i} = t_i$ and $T_{2i} > t_i$.  Therefore, the likelihood contribution becomes 
$$ 
P(t_i < T_{1i} < t_i+dt_i, T_{2i} > t_i|T_{1i} > \tau_{iL}, T_{2i} > \tau_{iL}) = 
\frac{\alpha \lambda_1 t_i^{\alpha-1} e^{-\lambda_1 t_i^{\alpha}} e^{-\lambda_2 t^{\alpha}}}{e^{-(\lambda_1+\lambda_2)\tau_{iL}^{\alpha}}} dt_i.
$$

\noindent Hence the likelihood function becomes 
\begin{eqnarray}
&L_1(\alpha, \lambda_1, \lambda_2) &= \prod_{i \in I_1}\bigg\{\alpha \lambda_1 t_i^{\alpha-1}e^{-(\lambda_1+\lambda_2)t_i^{\alpha}}\bigg\}^{\nu_i}\bigg\{
\frac{\alpha \lambda_1 t_i^{\alpha-1}e^{-(\lambda_1+\lambda_2)t_i^{\alpha}}}{e^{-(\lambda_1+\lambda_2)\tau_{iL}^{\alpha}}}\bigg\}^{1-\nu_i}\nonumber \\
&&\times \prod_{i \in I_2}\bigg\{\alpha \lambda_2 
t_i^{\alpha-1}e^{-(\lambda_1+\lambda_2)t_i^{\alpha}}\bigg\}^{\nu_i}\bigg\{
\frac{\alpha \lambda_2 t_i^{\alpha-1}e^{-(\lambda_1+\lambda_2)t_i^{\alpha}}}{e^{-(\lambda_1+\lambda_2)\tau_{iL}^{\alpha}}}\bigg\}^{1-\nu_i}\nonumber \\
&&\times \prod_{i \in I_0} \bigg\{e^{-(\lambda_1+\lambda_2)t_i^{\alpha}}\bigg\}^{\nu_i}\bigg\{\frac{e^{-(\lambda_1+\lambda_2)t_i^{\alpha}}}{e^{-(\lambda_1+\lambda_2)\tau_{iL}^{\alpha}}}\bigg\}^{1-\nu_i} \nonumber  \\
&  & = \alpha^{m} \lambda_1^{m_1} \lambda_2^{m_2} \prod_{i \in I_1 \cup I_2} t_i^{\alpha-1} \times
e^{-(\lambda_1+\lambda_2)\big[\sum_{i=1}^{n}t_i^{\alpha} - \sum_{i=1}^{n}(1-\nu_i)\tau_{iL}^{\alpha}\big]}.\nonumber
\end{eqnarray}
The log-likelihood can be written as
\begin{equation}
\log L_1(\alpha, \lambda_1, \lambda_2) = m \log \alpha + m_1 \log \lambda_1 + m_2 \log \lambda_2 + (\alpha - 1)w_1 - (\lambda_1+\lambda_2)w_2(\alpha),   \label{ll}
\end{equation}
where 
\begin{equation}
w_1 = \sum_{i \in I_1 \cup I_2} \log t_i \ \ \ \hbox{and} \ \ \ w_2(\alpha) = \sum_{i=1}^{n}t_i^{\alpha} - \sum_{i=1}^{n}(1-\nu_i)\tau_{iL}^{\alpha}.
\label{w1w2}
\end{equation}
For known $\alpha$, the MLEs of $\lambda_1$ and $\lambda_2$ can 
obtained by taking derivatives of \eqref{ll} with respect to $\lambda_1$ and $\lambda_2$, respectively, and
equating them to zero as;
$$
\widehat{\lambda}_1(\alpha) = \frac{m_1}{w_2(\alpha)} \qquad\text{ and }\qquad \widehat{\lambda}_2(\alpha) =
\frac{m_2}{w_2(\alpha)}.
$$
It easily follows from the second derivatives matrix of \eqref{ll} that for known $\alpha$, when $m_1 > 0$ and $m_2 > 0$, 
the MLEs of $\lambda_1$ and $\lambda_2$ exist and they are unique.  When $\alpha$ is unknown, 
putting back $\widehat{\lambda}_1(\alpha)$ and $\widehat{\lambda}_2(\alpha)$ in \eqref{ll}, we get the profile
log-likelihood for $\alpha$ (without the additive constant) as 
$$
p(\alpha) = m\log \alpha - m\log w_2(\alpha) + \alpha w_1.
$$ 
The MLE of $\alpha$, say $\widehat{\alpha}$, can be obtained by maximizing $p(\alpha)$ with respect to $\alpha$.  Once 
$\widehat{\alpha}$ is obtained, the MLEs of $\lambda_1$ and $\lambda_2$ can be obtained as $\displaystyle \widehat{\lambda}_1
= \widehat{\lambda}_1(\widehat{\alpha})$ and $\displaystyle \widehat{\lambda}_2 = \widehat{\lambda}_2(\widehat{\alpha})$, 
respectively.  The following result is useful for further development.  

\noindent {\sc Lemma 1}: For $m_1 > 0$,  $m_2 > 0$, and for a given $\alpha$, $\widehat{\lambda}_1(\alpha)$ and $\widehat{\lambda}_2(\alpha)$ are the unique MLEs of $\lambda_1$ and $\lambda_2$, respectively.   

\begin{proof} It is straightforward, and hence is omitted here. 
\end{proof}

\noindent {\sc Lemma 2}: Define
\begin{eqnarray}
& d(\alpha) & = \bigg(\sum_{i=1}^{n}t_i^{\alpha} - \sum_{i=1}^{n}(1-\nu_i)\tau_{iL}^{\alpha}\bigg)\times\bigg(\sum_{i=1}^{n}t_i^{\alpha}(\log t_i)^2 - \sum_{i=1}^{n}(1-\nu_i)\tau_{iL}^{\alpha}(\log \tau_{iL})^2\bigg) \nonumber \\
&& - \bigg(\sum_{i=1}^{n}t_i^{\alpha}\log t_i - \sum_{i=1}^{n}(1-\nu_i)\tau_{iL}^{\alpha}\log \tau_{iL}\bigg)^2. \nonumber
\end{eqnarray}
If for $\alpha > 0$, $d(\alpha) \geq 0$, then the function $p(\alpha)$ is unimodal.

\begin{proof} To show that $p(\alpha)$ is unimodal, first we shall show that $p(\alpha)$ is concave when the sufficient condition is satisfied. We have,
   $$p^{\prime\prime}(\alpha) = -m\bigg[\frac{1}{\alpha^2} + \frac{w_2(\alpha)w_2''(\alpha) -
(w_2'(\alpha))^2}{(w_2(\alpha))^2}\bigg].$$
Note that $d(\alpha)=w_2(\alpha)w_2''(\alpha) - (w_2'(\alpha))^2$.
Therefore, if for $\alpha > 0$, $d(\alpha) \geq 0$, we have $p''(\alpha) < 0$, and hence $p(\alpha)$ is concave. Now, using the fact
\[
\lim_{\alpha \to 0+} p(\alpha) = \lim_{\alpha \to \infty} p (\alpha)= -\infty,
\]
we conclude that $p(\alpha)$ is unimodal, provided $d(\alpha) \geq 0$. 
\end{proof}

Therefore, if the given data is such that $d(\alpha) \geq 0$, when $m_1 >0$ and $m_2 > 0$, we immediately obtain that the MLEs of $\alpha$, $\lambda_1$ and $\lambda_2$ exist and they are
unique.  Since $p(\alpha)$ is unimodal when the sufficient condition is satisfied, it is quite easy to maximize $p(\alpha)$. After checking the sufficient condition for the data, we can use the standard algorithm like Newton-Raphson method to maximize $p(\alpha)$. Alternatively, by equating $p'(\alpha)$ to zero we obtain the following fixed point equation:
\begin{equation}
\alpha = h(\alpha) =\frac{mw_2(\alpha)}{mw_2'(\alpha)-w_1w_2(\alpha)}.   \label{fpe}
\end{equation}
Clearly, $\widehat{\alpha}$ is a fixed point solution of \eqref{fpe}.  A very simple iterative procedure may be used to compute 
$\widehat{\alpha}$.  First, we start with an initial value of $\alpha$, say $\alpha^{(0)}$. Then, obtain  $\alpha^{(1)} = h(\alpha^{(0)})$.
Continue this process until convergence is achieved.  Once $\widehat{\alpha}$ is obtained, $\widehat{\lambda}_1$ and 
$\widehat{\lambda}_2$ can be easily obtained as described before.

\noindent {\sc Theorem 1}: For $m_1 > 0$ and $m_2 > 0$, $\widehat{\lambda}_1$, $\widehat{\lambda}_2$, and $\widehat{\alpha}$ are the unique MLEs of $\lambda_1$, $\lambda_2$ and $\alpha$, respectively, if $d(\alpha) \geq 0$ for $\alpha > 0$.

\begin{proof} Follows from Lemma 1 and Lemma 2.  
\end{proof}

Note that when the sufficient condition is not satisfied for any given left truncated right censored data, i.e., if $d(\alpha) < 0$ for some $\alpha > 0$, we cannot comment on the uniqueness of the MLEs. Also, note that although the MLEs can be calculated quite conveniently, the associated exact confidence intervals cannot be obtained.  Hence we propose to use the parametric percentile bootstrap and parametric biased corrected bootstrap method to compute the confidence intervals of the unknown parameters, as given below.

\subsection{\sc Bootstrap confidence intervals}
One can construct both parametric and non-parametric bootstrap confidence intervals in this situation.  However, as the data contains both truncation and censoring, a parametric bootstrap confidence interval is expected to be more efficient than a non-parametric one; Balakrishnan, Kundu, Ng and Kannan (2007) made a similar observation in the context of analysis of censored data from step-stress reliability experiments. Parametric bootstrap confidence intervals for the model parameters can be constructed in the following manner. 

After obtaining the MLEs  $\widehat{\alpha}$, $\widehat{\lambda}_1$ and $\widehat{\lambda}_2$ of the model parameters, using these estimates as the true values of the parameters, a sample of size $n$ can be obtained in the same sampling framework of competing risks with left truncation and right censoring. From this sample, one can obtain the MLEs of the parameters in the same way as described above, let these MLEs be denoted by $\widehat{\alpha}^*$, $\widehat{\lambda}_1^*$, and $\widehat{\lambda}_2^*$. This process is then repeated for $B$ times, to obtain $B$ such bootstrap samples. The MLEs of the parameters are obtained from each of these $B$ samples, that is, we now have the MLEs for the bootstrap samples as $(\widehat{\alpha}_1^*, \widehat{\lambda}_{11}^*, \widehat{\lambda}_{21}^*)$, $(\widehat{\alpha}_2^*, \widehat{\lambda}_{12}^*, \widehat{\lambda}_{22}^*)$,...,$(\widehat{\alpha}_B^*, \widehat{\lambda}_{1B}^*, \widehat{\lambda}_{2B}^*)$. Then, a $100(1-\beta)\%$ parametric bootstrap confidence interval for a model parameter, say $\lambda_1$ is calculated as 
\[
(\widehat{\lambda}_1 - b_{\lambda_1} - z_{\beta/2}\sqrt{v_{\lambda_1}}, \widehat{\lambda}_1 - b_{\lambda_1} + z_{\beta/2}\sqrt{v_{\lambda_1}}),
\]
where $b_{\lambda_1}$ and $v_{\lambda_1}$ are the bootstrap bias and bootstrap variance for the parameter $\lambda_1$, and $z_{\beta}$ is the upper $\beta$-percentage point of standard normal distribution. The bootstrap bias and variance are given by 
\[
b_{\lambda_1} = \overline{\widehat{\lambda}_{1}^*}-\widehat{\lambda}_1, \quad v_{\lambda_1} =
\frac{1}{B-1}\sum_{i=1}^{B}\left(\widehat{\lambda}_{1i}^* - \overline{\widehat{\lambda}_{1}^*}\right)^2,
\]
where $\overline{\widehat{\lambda}_{1}^*} = \frac{1}{B}\sum_{i=1}^{B}\widehat{\lambda}_{1i}^*$. The parametric bootstrap confidence intervals for $\alpha$ and $\lambda_2$ can be constructed in a similar way. 

Yet another type of bootstrap confidence intervals for the parameters may be obtained simply by choosing appropriate percentile points from the ordered values of the bootstrap estimates of the parameters. Thus, for example, for the parameter $\lambda_1$, a $100(1-\beta)\%$ bootstrap confidence interval can be given by $(\widehat{\lambda}_{1([B\beta/2])}^*, \widehat{\lambda}_{1([B(1-\beta/2)])}^*)$, where $\widehat{\lambda}_{1(1)}^*, \widehat{\lambda}_{1(2)}^*, ..., \widehat{\lambda}_{1(B)}^*$ are the ordered bootstrap estimates of the parameter $\lambda_1$, and $[x]$ indicates the greatest integer value of the number $x$.

\section{\sc Bayesian analysis}

In this section we consider the Bayesian inference of the unknown parameters.  First we consider the case when the common shape 
parameter is known and we obtain the Bayes estimates and the associated credible set of the scale parameters.  Then we consider the
case when the common shape parameter is also unknown.  In this case the Bayes estimates and the associated credible intervals cannot 
be obtained in explicit forms, and we use importance sampling technique to compute the Bayes estimates and the credible 
intervals.  In developing the Bayes estimates we have assumed the squared error loss function although any other loss function can be
easily incorporated.

\subsection{\sc Prior Assumptions}

Following Pena and Gupta (1990) we assume DG prior on the scale parameters $\lambda_1$ and $\lambda_2$, and they can be 
described as follows.   Assume that $\lambda = \lambda_1 + \lambda_2$ has
a gamma distribution with parameters $a_0$ and $b_0$, $a_0 > 0, b_0 > 0$, (denoted by GA$(a_0,b_0)$) and $p = \lambda_1/\lambda$ has a
beta distribution with parameters $a_1$ and $a_2$, $a_1 > 0, a_2 > 0$ (denoted by Beta$(a_1, a_2)$).  That is, $\lambda$ has the probability
density function (PDF) given by 
\[
\pi(\lambda|a_0, b_0) = \frac{b_0^{a_0}}{\Gamma(a_0)}\lambda^{a_0-1}e^{-b_0 \lambda}, \;\;\;\; \lambda > 0,
\]
and $p$ has the PDF given by 
\[
\pi(p|a_1, a_2) = \frac{\Gamma(a_1+a_2)}{\Gamma(a_1)\Gamma(a_2)} p^{a_1-1}(1-p)^{a_2-1}, \;\;\;\; p>0. 
\]
Then, the joint prior distribution of $\lambda_1$ and $\lambda_2$ can be obtained as 
\begin{equation}
\pi_1(\lambda_1, \lambda_2|a_0, b_0, a_1, a_2) = \frac{\Gamma(a_1+a_2)}{\Gamma(a_0)}(b_0 \lambda)^{a_0 - a_1 - a_2} \times
\frac{b_0^{a_1}}{\Gamma(a_1)}\lambda_1^{a_1-1}e^{-b_0 \lambda_1} \times \frac{b_0^{a_2}}{\Gamma(a_2)}\lambda_2^{a_2-1}e^{-b_0
\lambda_2},\nonumber
\end{equation}
with $\lambda_1, \lambda_2 > 0$, $\lambda = \lambda_1 + \lambda_2$.  This is known as Dirichlet-Gamma distribution, and we denote it by 
DG$(b_0, a_0, a_1, a_2)$. Using Theorem 2 of Pena and Gupta (1990), it can be very easily seen that 

\[
E(\lambda_i) = \frac{a_0a_i}{b_0(a_1+a_2)}, 
\] 
and 
\[
V(\lambda_i) = \frac{a_0a_i}{b_0^2(a_1+a_2)}\times \bigg\{\frac{(a_i+1)(a_0+1)}{a_1+a_2+1}-\frac{a_0a_i}{a_1+a_2}\bigg\}
\]
for $i=$1,2.

The joint prior of $\lambda_1$ and $\lambda_2$ is a conjugate prior, when $\alpha$ is known, and it is very flexible.  
The joint PDF can take variety of shapes and the dependency between $\lambda_1$ and $\lambda_2$ can be controlled through 
the hyper-parameters. For example, when $a_0 = a_1+a_2$, then $\lambda_1$ and $\lambda_2$ are independent.  Further, 
$\lambda_1$ and $\lambda_2$ are positively, or negatively correlated depending on whether $a_0 > a_1+a_2$, or $a_0 < a_1+a_2$, 
respectively.  Moreover, using the method suggested by Kundu and Pradhan (2011), the generation from a DG distribution can be 
performed very conveniently.

When the common shape parameter is also unknown, we need to assume some prior on $\alpha$.  In this case we do not make any 
specific prior assumption on $\alpha$.  When the shape parameter is also unknown, the joint conjugate priors do not exist.  In this 
case following the approach of Berger and Sun (1993) or Kundu (2008), it is assumed that the scale parameters $(\lambda_1, \lambda_2)$ 
has the same prior as described above, and no specific form on the prior $\pi_2(\alpha)$ on $\alpha$ is assumed here.  It is assumed
that $\alpha$ has log-concave PDF with support on $(0, \infty)$ and it is independent of $\lambda_1$ and $\lambda_2$.
\subsection{\sc Common shape parameter $\alpha$ is known}
In this case the posterior distribution of $\lambda_1$ and $\lambda_2$ becomes
\begin{eqnarray}
&\pi(\lambda_1, \lambda_2|\textrm{data}, \alpha, a_0, b_0, a_1, a_2) &\propto L_1(\alpha ,\lambda_1, \lambda_2) \times \pi(\lambda_1, \lambda_2|a_0, b_0, a_1, a_2) \nonumber \\
&& \propto \frac{\Gamma(a_1+m_1+a_2+m_2)}{\Gamma(a_0+m_1+m_2)}\bigg\{(b_0+w_2(\alpha))\lambda\bigg\}^{(a_0+m_1+m_2)-(a_1+m_1)-(a_2+m_2)} \nonumber \\
&& \times \frac{(b_0+w_2(\alpha))^{a_1+m_1}}{\Gamma(a_1+m_1)}\lambda_1^{a_1+m_1-1}e^{(-b_0+w_2(\alpha))\lambda_1} \nonumber \\
&& \times \frac{(b_0+w_2(\alpha))^{a_2+m_2}}{\Gamma(a_2+m_2)}\lambda_2^{a_2+m_2-1}e^{(-b_0+w_2(\alpha))\lambda_2}. \nonumber
\end{eqnarray}
Clearly 
\[
\pi(\lambda_1, \lambda_2|\textrm{data}, \alpha, a_0, b_0, a_1, a_2) \sim \textrm{DG}(b_0+w_2(\alpha), a_0+m_1+m_2, a_1+m_1, a_2+m_2).
\]
Therefore, the Bayes estimates for $\lambda_1$ and $\lambda_2$ with respect to squared error loss function become
\begin{eqnarray}
\widehat{\lambda}_1^B & = & E_{\textrm{posterior}}(\lambda_1) =
\frac{(a_0+m_1+m_2)(a_1+m_1)}{(b_0+w_2(\alpha))(a_1+m_1+a_2+m_2)},\nonumber\\
\widehat{\lambda}_2^B & = & E_{\textrm{posterior}}(\lambda_2) =
\frac{(a_0+m_1+m_2)(a_2+m_2)}{(b_0+w_2(\alpha))(a_1+m_1+a_2+m_2)},\nonumber
\end{eqnarray}
and the posterior variances are 
$$V_{(posterior)}(\lambda_1) = A_1 \times B_1,\qquad\text{and}\qquad V_{(posterior)}(\lambda_2) = A_2 \times
B_2,$$
where for $i = 1, 2$,
\begin{align*}
{} & A_i = \frac{(a_0+m_1+m_2)(a_i+m_i)}{(b_0+w_2(\alpha))^2(a_1+m_1+a_2+m_2)},\\
\shortintertext{and}
{} & B_i = \frac{(a_i+m_i+1)(a_0+m_1+m_2+1)}{a_1+m_1+a_2+m_2+1} -
\frac{(a_0+m_1+m_2)(a_i+m_i)}{a_1+m_1+a_2+m_2}.
\end{align*}
Now we describe how to construct a 100(1-$\gamma$)\% credible set of $(\lambda_1, \lambda_2)$.  Let us recall that a set 
$C_{\alpha, 1-\gamma}$ is said to be a 100(1-$\gamma$)\% credible set of $(\lambda_1, \lambda_2)$ if
$$
P((\lambda_1, \lambda_2) \in C_{\alpha, 1-\gamma}) = 
\int \int_{C_{\alpha, 1-\gamma}} \pi(\lambda_1, \lambda_2|\textrm{data}, \alpha, a_0, b_0, a_1, a_2) d\lambda_1 d \lambda_2 = 
1 - \gamma.
$$
Now using the fact that if $(\lambda_1, \lambda_2) \sim \textrm{DG}(b_0+w_2(\alpha), a_0+m_1+m_2, a_1+m_1, a_2+m_2)$, then 
$\lambda_1 + \lambda_2 \sim \hbox{GA}(a_0+m_1+m_2, b_0+w_2(\alpha))$ and $\displaystyle \frac{\lambda_1}{\lambda_1 + \lambda_2} 
\sim \hbox{Beta}(a_1+m_1, a_2+m_2)$, and they are independently distributed, we obtain
\begin{equation}
C_{\alpha, 1- \gamma} = \{(\lambda_1, \lambda_2); \lambda_1 > 0, \lambda_2 > 0, A \le \lambda_1 + \lambda_2 \le B, C \le 
\frac{\lambda_1}{\lambda_1+\lambda_2} \le D\}.\nonumber
\end{equation}
Here $A, B, C, D$ are such that
$$
P(A \le U \le B) = 1 - \gamma_1 \ \ \ \hbox{and} \ \ \ \ P(C \le V \le D) = 1 - \gamma_2,
$$
$U \sim \hbox{GA}(a_0+m_1+m_2, b_0+w_2(\alpha))$ and $V \sim \hbox{Beta}(a_1+m_1, a_2+m_2)$, and they are independently 
distributed.  Further, $\gamma_1$ and $\gamma_2$ are such that $1 - \gamma = (1-\gamma_1)(1-\gamma_2)$.  Note that $C_{\alpha, 1- \gamma}$
is a trapezoid enclosed by the following straight lines
$$
(i) \lambda_1 + \lambda_2 = A, \ \ (ii) \lambda_1 + \lambda_2 = B, \ \ (iii) \lambda_1(1-D) = \lambda_2 D, \ \ (iv) 
\lambda_1(1-C) = \lambda_2 C,
$$
and  the area of the credible set is $(B^2-A^2)(D-C)/2$.

\subsection{\sc Common shape parameter $\alpha$ is not known} 
In this case the joint posterior density of $\alpha$, $\lambda_1$ and $\lambda_2$ is given by 
\begin{equation}
\pi(\alpha, \lambda_1, \lambda_2|\textrm{data}) = \frac{L_1(\alpha, \lambda_1, \lambda_2)\pi_1(\lambda_1,
\lambda_2)\pi_2(\alpha)}{\int_0^{\infty}\int_0^{\infty}\int_0^{\infty}L_1(\alpha, \lambda_1, \lambda_2)\pi_1(\lambda_1,
\lambda_2)\pi_2(\alpha)d\alpha d\lambda_1 d\lambda_2}.\nonumber
\end{equation}
Therefore, the Bayes estimate of any function of $\alpha$, $\lambda_1$ and $\lambda_2$, say $g(\alpha,
\lambda_1, \lambda_2)$, with respect to squared error loss would be 
\begin{equation}
\widehat{g}_B(\alpha, \lambda_1, \lambda_2) = E_{\textrm{posterior}}(g(\alpha, \lambda_1, \lambda_2)).   \label{beg}
\end{equation}
It is clear that even if we know explicitly $\pi_2(\alpha)$, \eqref{beg} cannot be calculated explicitly for general 
$g(\alpha, \lambda_1, \lambda_2)$.  We need the following results for further development.  First note that the joint 
posterior distribution of $(\alpha, \lambda_1, \lambda_2)$ can be written as
$$
\pi(\alpha, \lambda_1, \lambda_2|data) = \pi(\alpha|data) \times \pi(\lambda_1, \lambda_2|data, \alpha),
$$
where the joint posterior distribution of $(\lambda_1, \lambda_2)$ given $\alpha$, $\pi(\lambda_1, \lambda_2|data, \alpha)$ is 
DG$(b_0+w_2(\alpha), a_0+m_1+m_2, a_1+m_1, a_2+m_2)$, and
\begin{equation}
\pi(\alpha|\textrm{data}) \sim \pi(\alpha)\alpha^m\prod_{i \in I_1\cup I_2}t_i^{\alpha} \times
\frac{1}{(b_0+w_2(\alpha))^{a_0+m}}\text{ for } \alpha>0.\nonumber
\end{equation}

   \noindent {\sc Lemma 3}: $\pi(\alpha|\textrm{data})$ is log-concave if
   $\widetilde{d}(\alpha)=\widetilde{w}_2^{\prime\prime}(\alpha)\widetilde{w}_2(\alpha)-\left(
   \widetilde{w}_2^\prime(\alpha) \right)^2\geq0$ for all $\alpha>0$, where $\widetilde{w}_2(\alpha)=b_0+w_2(\alpha)$.
\begin{proof}
   Note that, for some constant $C$,
   \[
   \log \pi(\alpha|\textrm{data}) = C + \log \pi(\alpha) + m\log \alpha + (\alpha-1)\sum_{i \in I_1\cup I_2}\log t_i -
   (a_0+m)\log \widetilde w_2(\alpha).
   \]
   Now, if $\widetilde d(\alpha)\geq0$ for all $\alpha>0$ and $\pi(\alpha)$ is log-concave, it follows immediately that
   $\pi(\alpha|\textrm{data})$ is also log-concave.
   \end{proof}

Kinderman and Monahan (1977) proposed generation of random variables using ratio of uniform random variables.  Devroye (1984) 
proposed a method to generate samples from a density function with log-concave PDF.  Once the samples from $\pi(\alpha|data)$ 
are drawn, the generation from $\pi(\lambda_1, \lambda_2|data, \alpha)$ from a DG distribution can be performed as suggested by
Kundu and Pradhan (2011).  We propose the following
algorithm to compute the Bayes estimates of $g(\alpha, \lambda_1, \lambda_2)$, and to construct associated
highest posterior density (HPD) credible interval.

\noindent {\sc Algorithm:} 
\begin{itemize}
	\item Step 1: Generate $\alpha$ from $\pi(\alpha|data)$ using the method proposed by Kinderman and Monahan (1977) or 
by Devroye (1984).
    \item  Step 2: For given $\alpha$, generate $(\lambda_1, \lambda_2)$ from $\pi(\lambda_1,
    \lambda_2|data,\alpha)$, using the method proposed by Kundu and Pradhan (2011).
    \item Step 3: Repeat steps 1 and 2 for $N$ times, and obtain copies of $(\alpha, \lambda_1, \lambda_2)$
    as $(\alpha_i, \lambda_{1i}, \lambda_{2i})$, and obtain $g_i = g(\alpha_i, \lambda_{1i}, \lambda_{2i})$ $i=1,...,N$.
	\item Step 4: The Bayes estimate of $g(\alpha, \lambda_1, \lambda_2)$ and the corresponding posterior variance can be 
obtained and calculate the Bayes estimates of the parameters, with respect to squared error loss function, as
$$
\widehat{g}_B(\alpha, \lambda_1, \lambda_2)  =  \frac{1}{N}\sum_{i=1}^N g_i  \ \ \ \hbox{and} \ \ \ 
\widehat{V}(g(\alpha, \lambda_1, \lambda_2))  =  \frac{1}{N} \sum_{i=1}^N (g_i - \widehat{g}_B(\alpha, \lambda_1, \lambda_2))^2,
$$
respectively.
    \item Step 6: To construct the HPD credible interval of $g(\alpha, \lambda_1, \lambda_2)$, first order $g_i$ as 
$g_{(1)} < g_{(2)} < \ldots < g_{(N)}$.  Then 100(1-2$\beta$)\% credible interval of $g(\alpha, \lambda_1, \lambda_2)$ becomes
$$(g_{(j)}, g_{(j+N-[2N\beta+1])}); \ \ \ \hbox{for} \ \ \ j = 1, \ldots, [2N\beta].$$
Therefore, 100(1-2$\beta$)\% HPD credible interval becomes $(g_{(j^*)}, g_{(j^*+N-2\beta)})$, where $j^*$ is such that
$$
g_{(j^*+N-[2N\beta]+1)} - g_{(j^*)} \le g_{(j+N-[2N\beta]+1)} - g_{(j)}, 
$$
for all $j = 1, \ldots, [2N\beta]$.
\end{itemize}

\section{\sc Simulation Study}
We compare the performances of the different methods proposed here by an extensive Monte Carlo simulation study. For the
simulation study, we have fixed 1980 as the left truncation year, and 1984 as the right censoring year. First of all, a certain
truncation percentage is fixed, to ensure the proportion on truncated observations in the data. Then the installation years of
machines are sampled from an arbitrary set of years. The installation years, arbitrarily, are divided into two parts: (1975 to
1979) and (1980 to 1983). Equal probabilities are attached to each of the installation years, that is, a probability of 0.2 is
attached to each of the years in the set 1975 to 1979, and a probability of 0.25 is attached to each of the years in the set 1980
to 1983. Then, for the specified proportion of truncated observations, installation years are sampled from these two sets using
with replacement sampling. 

The lifetimes of the machines are sampled from two independent Weibull distributions, which correspond to the two causes of
failure in this setup of competing risks. Then, for each unit, whichever of the two lifetimes is smaller, is added to the unit's
installation year, to get the year of its failure. At this point, the cause of failure of the unit, that is, which one of the two
randomly generated Weibull lifetimes is smaller, is also noted. 

Note that the year of left truncation is 1980. This means that any failure that might have occurred before 1980 would not be known
to us. Hence, if the year of failure of a machine turns out to be less than 1980, that unit is completely discarded, and for that
unit, installation year, lifetimes and hence failure year, are generated again. Finally, again without any loss of generality, we
fix 1984 as the right censoring year, that is, any unit that fails after 1984 is treated as a right censored unit. It is worthy of
mentioning here that throughout this process, we keep in mind that we should have sufficiently many censored observations in our
data, for the given parameterization of the Weibull distribution.    

For simulation, we choose two sets of model parameters as follows: $(\alpha, \lambda_1, \lambda_2)$ = (2, 0.0625, 0.04) and (0.5,
0.378, 0.408). To see the performance of the methods under different levels of truncation, we fix the truncation percentages at
$10\%$, and $30\%$. These choices, along with the chosen years of left truncation and right censoring, produce enough proportion
of censored observations, along with the desired truncation proportions. For the Bayesian inference it is assumed that
$\pi_2(\alpha) \sim$ GA$(c,d)$, and the hyper-parameters take the following values: $a_0 = a_1 = a_2 = b_0 = c = d$ = 0.0001.

In each case we compute the MLEs of the unknown parameters and the associated 95\% bias-corrected bootstrap (BC-bootstrap) and
percentile bootstrap (P-bootstrap) confidence intervals.  We report the average bias, root mean square error (RMSE) of the MLEs,
the average confidence lengths (AL) and the coverage percentages (CP) over 1000 replications.  We also compute the Bayes estimates
and the associated symmetric and HPD credible intervals of the unknown parameters based on the above priors and the corresponding
hyper-parameters.  In this case also we report the average bias, RMSE of the Bayes estimates, the average credible lengths (AL)
and the coverage percentages (CP) over 1000 replications.  All the results are reported in Tables 1- 8.    

\begin{table}[p]
\caption{Performance of the MLE and CI for the model parameters $(\alpha, \lambda_1, \lambda_2)$ = (2, 0.0625, 0.04)}
\begin{center}
\begin{tabular}{*{9}{c}}
\toprule
\multicolumn{9}{c}{$n$ = 100}\\
\midrule
 & & & & & \multicolumn{2}{c}{BC-boot} & \multicolumn{2}{c}{P-boot} \\ 
\cmidrule(lr){6-7}\cmidrule(lr){8-9}
Parameter & Trunc. & Bias & RMSE & Nominal CL & CP & AL & CP & AL\\
\midrule
\multirow{4}{*}{$\alpha$}    & 10\% &  0.214 & 1.059 & 90\% & 0.884 & 0.880 & 0.868 & 0.859\\
& & &                                                & 95\% & 0.927 & 1.048 & 0.908 & 1.022\\
                             & 30\% & 0.050  & 0.325 & 90\% & 0.915 & 0.701 & 0.889 & 0.693\\
& & &                                                & 95\% & 0.952 & 0.835 & 0.938 & 0.819\\
\midrule                                                                        
\multirow{4}{*}{$\lambda_1$} & 10\% & -0.002 & 0.023 & 90\% & 0.843 & 0.058 & 0.858 & 0.057\\
& & &                                                & 95\% & 0.895 & 0.069 & 0.902 & 0.067\\
                             & 30\% & 0.001 & 0.020 & 90\% & 0.873 & 0.063 & 0.881 & 0.061\\
& & &                                                & 95\% & 0.915 & 0.075 & 0.923 & 0.073\\
\midrule                                                                        
\multirow{4}{*}{$\lambda_2$} & 10\% & -0.001 & 0.015 & 90\% & 0.848 & 0.041 & 0.856 & 0.040\\
             & & &                                   & 95\% & 0.887 & 0.049 & 0.899 & 0.047\\
                             & 30\% & -0.000 & 0.013 & 90\% & 0.878 & 0.043 & 0.885 & 0.042\\
& & &                                                & 95\% & 0.922 & 0.051 & 0.934 & 0.049\\
\bottomrule
\end{tabular}
\end{center}
\end{table}

\begin{table}[p]
\caption{Performance of the MLE and CI for the model parameters $(\alpha, \lambda_1, \lambda_2)$ = (2, 0.0625, 0.04)}
\begin{center}
\begin{tabular}{*{9}{c}}
\toprule
\multicolumn{9}{c}{$n$ = 200}\\
\midrule
 & & & & & \multicolumn{2}{c}{BC-boot} & \multicolumn{2}{c}{P-boot} \\ 
\cmidrule(lr){6-7}\cmidrule(lr){8-9}
Parameter & Trunc. & Bias & RMSE & Nominal CL & CP & AL & CP & AL\\
\midrule
\multirow{4}{*}{$\alpha$}    & 10\% & 0.050 & 0.392 & 90\% & 0.883 & 0.522 & 0.869 & 0.518\\
& & &                                                & 95\% & 0.944 & 0.622 & 0.924 & 0.613\\
                             & 30\% & 0.019 & 0.150 & 90\% & 0.923 & 0.475 & 0.907 & 0.472\\
& & &                                                & 95\% & 0.962 & 0.566 & 0.955 & 0.559\\
\midrule                                                                        
\multirow{4}{*}{$\lambda_1$} & 10\% & -0.001 & 0.015 & 90\% & 0.858 & 0.042 & 0.860 & 0.041\\
& & &                                                & 95\% & 0.914 & 0.050 & 0.917 & 0.049\\
                             & 30\% & -0.000 & 0.013  & 90\% & 0.894 & 0.044 & 0.899 & 0.043\\
& & &                                                & 95\% & 0.941 & 0.052 & 0.940 & 0.051\\
\midrule                                                                        
\multirow{4}{*}{$\lambda_2$} & 10\% & -0.000 & 0.010 & 90\% & 0.877 & 0.030 & 0.885 & 0.029\\
& & &                                                & 95\% & 0.925 & 0.035 & 0.937 & 0.034\\
                             & 30\% & -0.000 & 0.009 & 90\% & 0.889 & 0.031 & 0.884 & 0.030\\
& & &                                                & 95\% & 0.931 & 0.036 & 0.931 & 0.036\\
\bottomrule
\end{tabular}
\end{center}
\end{table}

\begin{table}[p]
\caption{Performance of the MLE and CI for the model parameters $(\alpha, \lambda_1, \lambda_2)$ = (0.5, 0.378, 0.408)}
\begin{center}
\begin{tabular}{*{9}{c}}
\toprule
\multicolumn{9}{c}{$n$ = 100}\\
\midrule
 & & & & & \multicolumn{2}{c}{BC-boot} & \multicolumn{2}{c}{P-boot} \\ 
\cmidrule(lr){6-7}\cmidrule(lr){8-9}
Parameter & Trunc. & Bias & RMSE & Nominal CL & CP & AL & CP & AL\\
\midrule
\multirow{4}{*}{$\alpha$}    & 10\% & 0.004 & 0.054 & 90\% & 0.892 & 0.176 & 0.883 & 0.175\\
& & &                                                & 95\% & 0.938 & 0.210 & 0.938 & 0.206\\
                             & 30\% & 0.006 & 0.056  & 90\% & 0.900 & 0.178 & 0.883 & 0.176\\
& & &                                                & 95\% & 0.942 & 0.212 & 0.938 & 0.208\\
\midrule                                                                        
\multirow{4}{*}{$\lambda_1$} & 10\% & -0.001 & 0.069 & 90\% & 0.896 & 0.222 & 0.882 & 0.220\\
& & &                                                & 95\% & 0.935 & 0.265 & 0.929 & 0.260\\
                             & 30\% & -0.002 & 0.073 & 90\% & 0.870 & 0.233 & 0.873 & 0.231\\
& & &                                                & 95\% & 0.925 & 0.277 & 0.924 & 0.273\\
\midrule                                                                        
\multirow{4}{*}{$\lambda_2$} & 10\% & -0.000 & 0.071 & 90\% & 0.890 & 0.231 & 0.882 & 0.229\\
& & &                                                & 95\% & 0.938 & 0.275 & 0.929 & 0.270\\
                             & 30\% & 0.002  & 0.073 & 90\% & 0.899 & 0.244 & 0.893 & 0.242\\
& & &                                                & 95\% & 0.955 & 0.291 & 0.950 & 0.286\\
\bottomrule
\end{tabular}
\end{center}
\end{table}

\begin{table}[p]
\caption{Performance of the MLE and CI for the model parameters $(\alpha, \lambda_1, \lambda_2)$ = (0.5, 0.378, 0.408)}
\begin{center}
\begin{tabular}{*{9}{c}}
\toprule
\multicolumn{9}{c}{$n$ = 200}\\
\midrule
 & & & & & \multicolumn{2}{c}{BC-boot} & \multicolumn{2}{c}{P-boot} \\ 
\cmidrule(lr){6-7}\cmidrule(lr){8-9}
Parameter & Trunc. & Bias & RMSE & Nominal CL & CP & AL & CP & AL\\
\midrule
\multirow{4}{*}{$\alpha$}    & 10\% &  0.003 & 0.037 & 90\% & 0.911 & 0.123 & 0.903 & 0.122\\
& & &                                                & 95\% & 0.949 & 0.146 & 0.942 & 0.144\\
                             & 30\% &  0.003 & 0.037 & 90\% & 0.902 & 0.123 & 0.899 & 0.122\\
& & &                                                & 95\% & 0.947 & 0.147 & 0.947 & 0.144\\
\midrule                                                                        
\multirow{4}{*}{$\lambda_1$} & 10\% &  0.002 & 0.047 & 90\% & 0.903 & 0.157 & 0.892 & 0.156\\
& & &                                                & 95\% & 0.943 & 0.187 & 0.934 & 0.185\\
                             & 30\% &  0.001 & 0.047 & 90\% & 0.917 & 0.165 & 0.911 & 0.164\\
& & &                                                & 95\% & 0.958 & 0.196 & 0.950 & 0.193\\
\midrule                                                                        
\multirow{4}{*}{$\lambda_2$} & 10\% &  0.000 & 0.047 & 90\% & 0.915 & 0.163 & 0.917 & 0.162\\
& & &                                                & 95\% & 0.953 & 0.194 & 0.949 & 0.192\\
                             & 30\% & -0.002 & 0.052 & 90\% & 0.908 & 0.172 & 0.906 & 0.171\\
& & &                                                & 95\% & 0.941 & 0.204 & 0.946 & 0.202\\
\bottomrule
\end{tabular}
\end{center}
\end{table}

\begin{table}[p]
\caption{Performance of BE and CRI for the model parameters $(\alpha, \lambda_1, \lambda_2)$ = (2, 0.0625, 0.04)}
\begin{center}
\begin{tabular}{*{9}{c}}
\toprule
\multicolumn{9}{c}{$n$ = 100}\\
\midrule
 & & & & & \multicolumn{2}{c}{Symm CRI} & \multicolumn{2}{c}{HPD CRI} \\ 
\cmidrule(lr){6-7}\cmidrule(lr){8-9}
Parameter & Trunc. & Bias & RMSE & Nominal CL & CP & AL & CP & AL\\
\midrule
\multirow{4}{*}{$\alpha$}    & 10\% & 0.037 & 0.230 & 90\% & 0.88 & 0.717 & 0.88 & 0.713\\
& & &                                               & 95\% & 0.93 & 0.854 & 0.93 & 0.849\\
                             & 30\% & 0.028 & 0.204 & 90\% & 0.90 & 0.662 & 0.90 & 0.659\\
& & &                                               & 95\% & 0.95 & 0.789 & 0.95 & 0.784\\
\midrule                                                     
\multirow{4}{*}{$\lambda_1$} & 10\% & 0.001 & 0.019 & 90\% & 0.91 & 0.060 & 0.88 & 0.058\\
& & &                                               & 95\% & 0.95 & 0.072 & 0.94 & 0.070\\
                             & 30\% & 0.003 & 0.019 & 90\% & 0.91 & 0.064 & 0.88 & 0.061\\
& & &                                               & 95\% & 0.95 & 0.076 & 0.95 & 0.074\\
\midrule                                                     
\multirow{4}{*}{$\lambda_2$} & 10\% & 0.001 & 0.013 & 90\% & 0.89 & 0.042 & 0.87 & 0.041\\
& & &                                               & 95\% & 0.94 & 0.051 & 0.94 & 0.049\\
                             & 30\% & 0.002 & 0.013 & 90\% & 0.90 & 0.044 & 0.88 & 0.042\\
& & &                                               & 95\% & 0.95 & 0.053 & 0.94 & 0.051\\
\bottomrule
\end{tabular}
\end{center}
\end{table}

\begin{table}[p]
\caption{Performance of BE and CRI for the model parameters $(\alpha, \lambda_1, \lambda_2)$ = (2, 0.0625, 0.04)}
\begin{center}
\begin{tabular}{*{9}{c}}
\toprule
\multicolumn{9}{c}{$n$ = 200}\\
\midrule
 & & & & & \multicolumn{2}{c}{Symm CRI} & \multicolumn{2}{c}{HPD CRI} \\ 
\cmidrule(lr){6-7}\cmidrule(lr){8-9}
Parameter & Trunc. & Bias & RMSE & Nominal CL & CP & AL & CP & AL\\
\midrule
\multirow{4}{*}{$\alpha$}    & 10\% & 0.015 & 0.156 & 90\% & 0.89 & 0.498 & 0.89 & 0.496\\
& & &                                               & 95\% & 0.95 & 0.593 & 0.95 & 0.591\\
                             & 30\% & 0.019 & 0.146 & 90\% & 0.90 & 0.465 & 0.90 & 0.463\\
& & &                                               & 95\% & 0.96 & 0.554 & 0.96 & 0.552\\
\midrule                                                     
\multirow{4}{*}{$\lambda_1$} & 10\% & 0.001 & 0.014 & 90\% & 0.89 & 0.043 & 0.88 & 0.042\\
& & &                                               & 95\% & 0.94 & 0.051 & 0.94 & 0.050\\
                             & 30\% & 0.001 & 0.014 & 90\% & 0.90 & 0.044 & 0.89 & 0.044\\
& & &                                               & 95\% & 0.95 & 0.053 & 0.94 & 0.052\\
\midrule                                                     
\multirow{4}{*}{$\lambda_2$} & 10\% & 0.001 & 0.009 & 90\% & 0.90 & 0.030 & 0.88 & 0.029\\
& & &                                               & 95\% & 0.95 & 0.036 & 0.93 & 0.035\\
                             & 30\% & 0.000 & 0.009 & 90\% & 0.90 & 0.030 & 0.88 & 0.030\\
& & &                                               & 95\% & 0.95 & 0.036 & 0.94 & 0.036\\
\bottomrule
\end{tabular}
\end{center}
\end{table}

\begin{table}[p]
\caption{Performance of BE and CRI for the model parameters $(\alpha, \lambda_1, \lambda_2)$ = (0.5,
0.378,0.408)}
\begin{center}
\begin{tabular}{*{9}{c}}
\toprule
\multicolumn{9}{c}{$n$ = 100}\\
\midrule
 & & & & & \multicolumn{2}{c}{Symm CRI} & \multicolumn{2}{c}{HPD CRI} \\ 
\cmidrule(lr){6-7}\cmidrule(lr){8-9}
Parameter & Trunc. & Bias & RMSE & Nominal CL & CP & AL & CP & AL\\
\midrule
\multirow{4}{*}{$\alpha$}    & 10\% &  0.007 & 0.055 & 90\% & 0.89 & 0.173 & 0.89 & 0.172\\
& & &                                                & 95\% & 0.94 & 0.206 & 0.94 & 0.205\\
                             & 30\% &  0.008 & 0.056 & 90\% & 0.90 & 0.174 & 0.89 & 0.173\\
& & &                                                & 95\% & 0.95 & 0.207 & 0.95 & 0.206\\
\midrule                                                      
\multirow{4}{*}{$\lambda_1$} & 10\% & -0.001 & 0.068 & 90\% & 0.89 & 0.219 & 0.87 & 0.216\\
& & &                                                & 95\% & 0.94 & 0.261 & 0.94 & 0.258\\
                             & 30\% &  0.003 & 0.071 & 90\% & 0.90 & 0.232 & 0.89 & 0.229\\
& & &                                                & 95\% & 0.95 & 0.276 & 0.94 & 0.273\\
\midrule                                                      
\multirow{4}{*}{$\lambda_2$} & 10\% & -0.001 & 0.071 & 90\% & 0.89 & 0.228 & 0.89 & 0.225\\
& & &                                                & 95\% & 0.95 & 0.271 & 0.95 & 0.268\\
                             & 30\% &  0.000 & 0.074 & 90\% & 0.90 & 0.241 & 0.88 & 0.238\\
& & &                                                & 95\% & 0.94 & 0.287 & 0.94 & 0.284\\
\bottomrule
\end{tabular}
\end{center}
\end{table}

\begin{table}[p]
\caption{Performance of BE and CRI for the model parameters $(\alpha, \lambda_1, \lambda_2)$ = (0.5,
0.378,0.408)}
\begin{center}
\begin{tabular}{*{9}{c}}
\toprule
\multicolumn{9}{c}{$n$ = 200}\\
\midrule
 & & & & & \multicolumn{2}{c}{Symm CRI} & \multicolumn{2}{c}{HPD CRI} \\ 
\cmidrule(lr){6-7}\cmidrule(lr){8-9}
Parameter & Trunc. & Bias & RMSE & Nominal CL & CP & AL & CP & AL\\
\midrule
\multirow{4}{*}{$\alpha$}    & 10\% &  0.003 & 0.037 & 90\% & 0.90 & 0.121 & 0.89 & 0.121\\
& & &                                                & 95\% & 0.95 & 0.144 & 0.95 & 0.144\\
                             & 30\% &  0.003 & 0.036 & 90\% & 0.91 & 0.122 & 0.91 & 0.121\\
& & &                                                & 95\% & 0.96 & 0.145 & 0.96 & 0.145\\
\midrule                                                      
\multirow{4}{*}{$\lambda_1$} & 10\% &  0.002 & 0.046 & 90\% & 0.92 & 0.156 & 0.92 & 0.155\\
& & &                                                & 95\% & 0.96 & 0.186 & 0.96 & 0.185\\
                             & 30\% & -0.000 & 0.052 & 90\% & 0.89 & 0.163 & 0.87 & 0.162\\
& & &                                                & 95\% & 0.94 & 0.195 & 0.94 & 0.193\\
\midrule                                                      
\multirow{4}{*}{$\lambda_2$} & 10\% &  0.001 & 0.050 & 90\% & 0.89 & 0.162 & 0.89 & 0.161\\
& & &                                                & 95\% & 0.95 & 0.193 & 0.95 & 0.192\\
                             & 30\% &  0.001 & 0.053 & 90\% & 0.88 & 0.171 & 0.88 & 0.170\\
& & &                                                & 95\% & 0.95 & 0.204 & 0.94 & 0.202\\
\bottomrule
\end{tabular}
\end{center}
\end{table}

Some of the points are quite clear from the Tables 1 - 8.  First of all it is observed in all the cases and for both the
approaches that as sample size increases, the bias and RMSE for all the parameters decrease.  It indicates the consistency
properties of the MLEs and the Bayes estimates.  It is observed that the truncation percentage has more effect on the performance
of the estimates of $\alpha$ than on $\lambda_1$ and $\lambda_2$ in most of the cases considered here.  It is observed that both
the bootstrap methods and both the credible intervals are quite satisfactory.  In most of the cases the coverage percentages are
very close to the corresponding nominal levels.  Another point is worth mentioning here that for the first set of parameter values
$(\alpha = 2.0, \lambda_1 = 0.0625, \lambda_2 = 0.04)$, it is observed that the bias and MSEs for 30\% truncation is smaller than
those of 10\% truncation.  It is mainly due to the design of the experiment.  It is observed in this case that for 10\% truncation
around 50\% data are censored, on the other hand for 30\% truncation around 40\% data are censored.  Therefore, in this case for
30\% truncation we have more complete observations than 10\% truncation, hence they provide better estimates.  Where as, for the
second set of parameter values $(\alpha = 0.5, \lambda_1 = 0.378, \lambda_2 = 0.408)$, in case of 10\% truncation around 33\% data
are censored, and for 30\% truncation around 35\% data are censored.  In this case it is observed that bias and MSEs at the
truncation levels are very close to each other in most of the cases considered.

Now comparing the performances between the MLEs and Bayes estimates it is quite clear that when $n$ = 100, the Bayes estimates
with non-informative priors provide better results than the MLEs in terms of lower biases and RMSE.  Also comparing the
performances between the confidence intervals and the credible intervals for $n$ = 100, it is quite apparent that the average
lengths of the HPD credible intervals are shorter than the symmetric credible intervals and also the two bootstrap confidence
intervals.  Moreover, it maintains the required coverage percentages also in all the cases.  Although, for $n$ = 200, the MLEs and
the Bayes estimators behave in a very similar manner in all respects.  Therefore, we propose to use the Bayes estimates with
non-informative priors and HPD credible intervals to analyze left truncated right censored competing risks data for moderate or
large sample sizes, for very large sample sizes it does not make any difference.

\section{\sc Illustrative Example}

In this section we provide the analysis of a data set for illustrative purposes. The data set is presented in the Appendix and it
is of size 100.  The truncation percentage is fixed at 30.  We note that 53 units are censored in this data set, and the number of
failures from Cause 1 and Cause 2 are 14, and 33, respectively.  We re-scale the data by dividing all the lifetimes by 100, mainly
for computational purposes. It is not going to affect in the inference procedure.

\begin{figure}[!ht]
	\centering
	\includegraphics[width=4in]{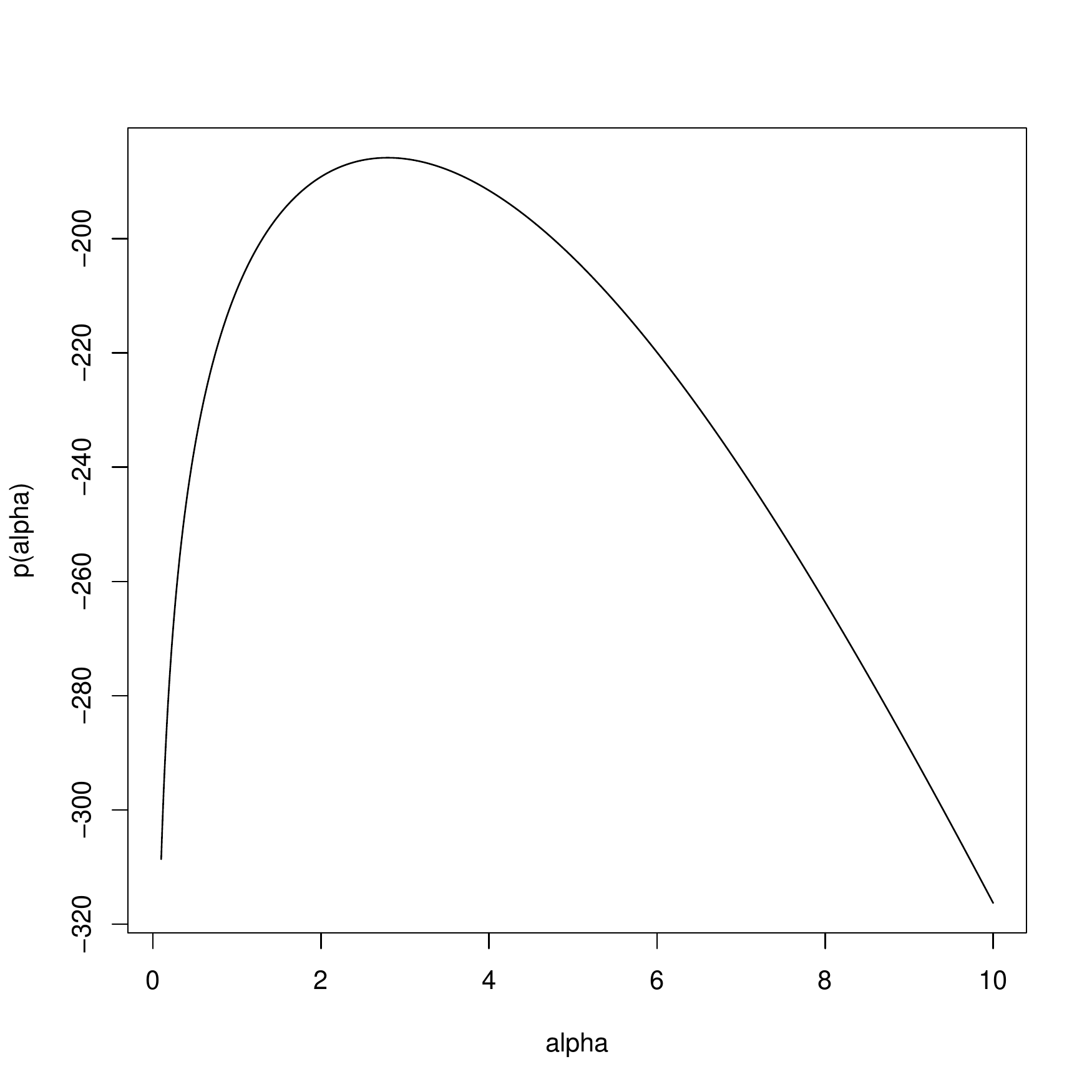}
	\caption{Profile-likelihood of $\alpha$}
\end{figure}

\begin{figure}[!ht]
   \centering
   \input{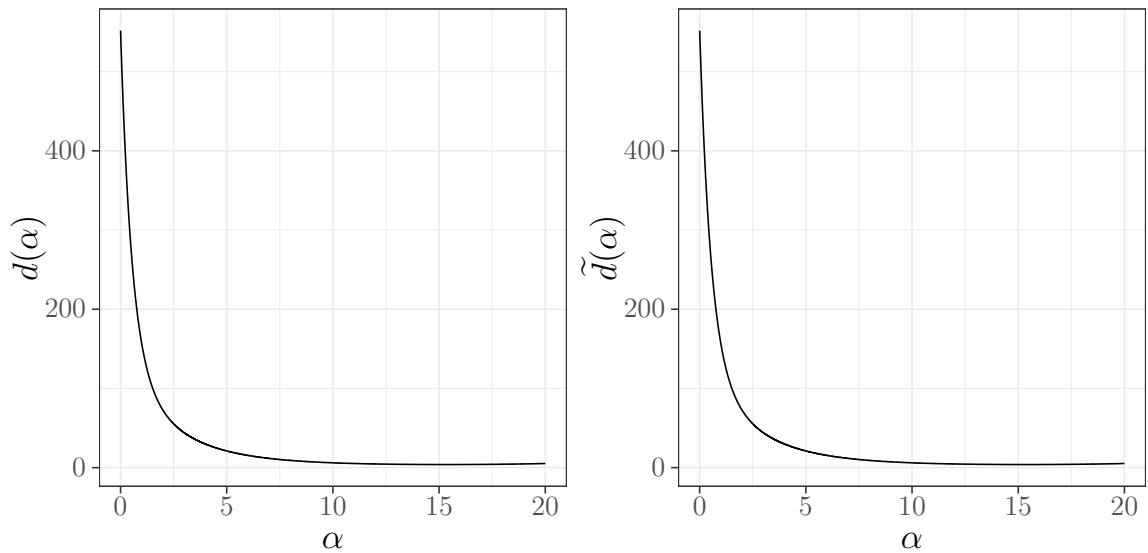}
   \caption{Plot of $d(\alpha)$ and $\widetilde d(\alpha)$}
\end{figure}

Before progressing further first we obtain the plot of $p(\alpha),\,d(\alpha)$, and $\widetilde d(\alpha)$ for this data. The plots are provided in Figures 1 and 2. Form Figure~1, it is clear that the profile log-likelihood function of $\alpha$ is a unimodal function with mode lying between 2 and 4. Figure 2 shows that $d(\alpha)$ and $\widetilde d(\alpha)$ are non-negative. Note that as $b_0=0.0001$, $d(\alpha)$ and $\widetilde d(\alpha)$ are very close.
The maximum likelihood estimates of the different parameters are as follows: $\widehat{\alpha}$ = 2.795, $\widehat{\lambda}_1$ = 6.759, and $\widehat{\lambda}_2$ = 15.932.  The associated bootstrap confidence intervals are also obtained and they are provided in Table \ref{data-ci}. We further obtain the Bayes estimates and the associated credible intervals of the unknown parameters based on the same set of priors as defined in the previous section.  We provide the plots of the histograms of the posterior samples on $\alpha$, $\lambda_1$ and $\lambda_2$ in Figure 3.  It is clear that the posterior distribution of $\alpha$ is quite
symmetric, whereas the posterior distributions of $\lambda_1$ and $\lambda_2$ are skewed for this data set.  The Bayes
estimates of $\alpha$, $\lambda_1$, and $\lambda_2$ with respect to squared error loss function are 2.781, 7.140, and
16.792, respectively.   The associated credible intervals are also provided in Table \ref{data-ci}.  It is clear that
the Bayes estimates and the MLEs are quite close to each other although the length of the HPD credible intervals are
slightly smaller than the bootstrap confidence intervals.

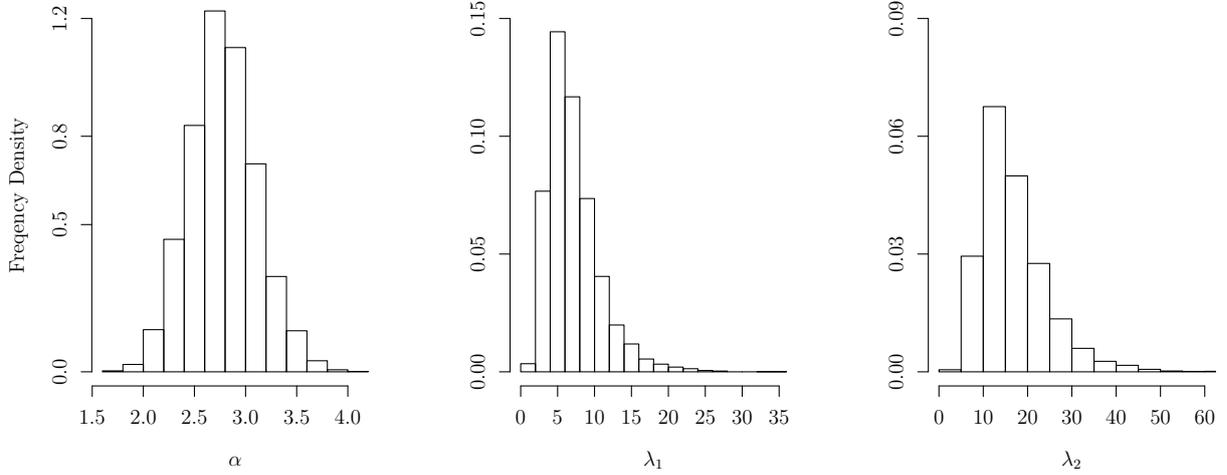
\begin{figure}[!ht]
\begin{center}
\begin{tikzpicture}[x=1pt,y=1pt]
\definecolor{fillColor}{RGB}{255,255,255}
\path[use as bounding box,fill=fillColor,fill opacity=0.00] (0,0) rectangle (469.75,216.81);
\begin{scope}
\path[clip] (  0.00,  0.00) rectangle (156.58,216.81);
\definecolor{drawColor}{RGB}{0,0,0}

\node[text=drawColor,anchor=base,inner sep=0pt, outer sep=0pt, scale=  0.66] at ( 86.21, 10.30) {$\alpha$};

\node[text=drawColor,rotate= 90.00,anchor=base,inner sep=0pt, outer sep=0pt, scale=  0.66] at (  7.13,112.36) {Freqency Density};
\end{scope}
\begin{scope}
\path[clip] ( 32.47, 40.39) rectangle (139.95,184.34);
\definecolor{drawColor}{RGB}{0,0,0}

\path[draw=drawColor,line width= 0.4pt,line join=round,line cap=round] ( 36.45, 45.72) rectangle ( 44.11, 46.07);

\path[draw=drawColor,line width= 0.4pt,line join=round,line cap=round] ( 44.11, 45.72) rectangle ( 51.76, 48.50);

\path[draw=drawColor,line width= 0.4pt,line join=round,line cap=round] ( 51.76, 45.72) rectangle ( 59.42, 61.62);

\path[draw=drawColor,line width= 0.4pt,line join=round,line cap=round] ( 59.42, 45.72) rectangle ( 67.07, 95.70);

\path[draw=drawColor,line width= 0.4pt,line join=round,line cap=round] ( 67.07, 45.72) rectangle ( 74.73,138.67);

\path[draw=drawColor,line width= 0.4pt,line join=round,line cap=round] ( 74.73, 45.72) rectangle ( 82.38,181.85);

\path[draw=drawColor,line width= 0.4pt,line join=round,line cap=round] ( 82.38, 45.72) rectangle ( 90.04,168.04);

\path[draw=drawColor,line width= 0.4pt,line join=round,line cap=round] ( 90.04, 45.72) rectangle ( 97.70,124.17);

\path[draw=drawColor,line width= 0.4pt,line join=round,line cap=round] ( 97.70, 45.72) rectangle (105.35, 81.68);

\path[draw=drawColor,line width= 0.4pt,line join=round,line cap=round] (105.35, 45.72) rectangle (113.01, 61.20);

\path[draw=drawColor,line width= 0.4pt,line join=round,line cap=round] (113.01, 45.72) rectangle (120.66, 49.89);

\path[draw=drawColor,line width= 0.4pt,line join=round,line cap=round] (120.66, 45.72) rectangle (128.32, 46.49);

\path[draw=drawColor,line width= 0.4pt,line join=round,line cap=round] (128.32, 45.72) rectangle (135.97, 45.86);
\end{scope}
\begin{scope}
\path[clip] (  0.00,  0.00) rectangle (469.75,216.81);
\definecolor{drawColor}{RGB}{0,0,0}

\path[draw=drawColor,line width= 0.4pt,line join=round,line cap=round] ( 32.63, 40.39) -- (128.32, 40.39);

\path[draw=drawColor,line width= 0.4pt,line join=round,line cap=round] ( 32.63, 40.39) -- ( 32.63, 36.43);

\path[draw=drawColor,line width= 0.4pt,line join=round,line cap=round] ( 51.76, 40.39) -- ( 51.76, 36.43);

\path[draw=drawColor,line width= 0.4pt,line join=round,line cap=round] ( 70.90, 40.39) -- ( 70.90, 36.43);

\path[draw=drawColor,line width= 0.4pt,line join=round,line cap=round] ( 90.04, 40.39) -- ( 90.04, 36.43);

\path[draw=drawColor,line width= 0.4pt,line join=round,line cap=round] (109.18, 40.39) -- (109.18, 36.43);

\path[draw=drawColor,line width= 0.4pt,line join=round,line cap=round] (128.32, 40.39) -- (128.32, 36.43);

\node[text=drawColor,anchor=base,inner sep=0pt, outer sep=0pt, scale=  0.66] at ( 32.63, 26.14) {1.5};

\node[text=drawColor,anchor=base,inner sep=0pt, outer sep=0pt, scale=  0.66] at ( 51.76, 26.14) {2.0};

\node[text=drawColor,anchor=base,inner sep=0pt, outer sep=0pt, scale=  0.66] at ( 70.90, 26.14) {2.5};

\node[text=drawColor,anchor=base,inner sep=0pt, outer sep=0pt, scale=  0.66] at ( 90.04, 26.14) {3.0};

\node[text=drawColor,anchor=base,inner sep=0pt, outer sep=0pt, scale=  0.66] at (109.18, 26.14) {3.5};

\node[text=drawColor,anchor=base,inner sep=0pt, outer sep=0pt, scale=  0.66] at (128.32, 26.14) {4.0};

\path[draw=drawColor,line width= 0.4pt,line join=round,line cap=round] ( 32.47, 45.72) -- ( 32.47,179.01);

\path[draw=drawColor,line width= 0.4pt,line join=round,line cap=round] ( 32.47, 45.72) -- ( 28.51, 45.72);

\path[draw=drawColor,line width= 0.4pt,line join=round,line cap=round] ( 32.47,101.26) -- ( 28.51,101.26);

\path[draw=drawColor,line width= 0.4pt,line join=round,line cap=round] ( 32.47,134.58) -- ( 28.51,134.58);

\path[draw=drawColor,line width= 0.4pt,line join=round,line cap=round] ( 32.47,179.01) -- ( 28.51,179.01);

\node[text=drawColor,rotate= 90.00,anchor=base,inner sep=0pt, outer sep=0pt, scale=  0.66] at ( 22.97, 45.72) {0.0};

\node[text=drawColor,rotate= 90.00,anchor=base,inner sep=0pt, outer sep=0pt, scale=  0.66] at ( 22.97,101.26) {0.5};

\node[text=drawColor,rotate= 90.00,anchor=base,inner sep=0pt, outer sep=0pt, scale=  0.66] at ( 22.97,134.58) {0.8};

\node[text=drawColor,rotate= 90.00,anchor=base,inner sep=0pt, outer sep=0pt, scale=  0.66] at ( 22.97,179.01) {1.2};
\end{scope}
\begin{scope}
\path[clip] (156.58,  0.00) rectangle (313.17,216.81);
\definecolor{drawColor}{RGB}{0,0,0}

\node[text=drawColor,anchor=base,inner sep=0pt, outer sep=0pt, scale=  0.66] at (242.80, 10.30) {$\lambda_1$};
\end{scope}
\begin{scope}
\path[clip] (189.06, 40.39) rectangle (296.54,184.34);
\definecolor{drawColor}{RGB}{0,0,0}

\path[draw=drawColor,line width= 0.4pt,line join=round,line cap=round] (193.04, 45.72) rectangle (198.57, 48.78);

\path[draw=drawColor,line width= 0.4pt,line join=round,line cap=round] (198.57, 45.72) rectangle (204.10,113.86);

\path[draw=drawColor,line width= 0.4pt,line join=round,line cap=round] (204.10, 45.72) rectangle (209.62,174.01);

\path[draw=drawColor,line width= 0.4pt,line join=round,line cap=round] (209.62, 45.72) rectangle (215.15,149.46);

\path[draw=drawColor,line width= 0.4pt,line join=round,line cap=round] (215.15, 45.72) rectangle (220.68,111.09);

\path[draw=drawColor,line width= 0.4pt,line join=round,line cap=round] (220.68, 45.72) rectangle (226.21, 81.71);

\path[draw=drawColor,line width= 0.4pt,line join=round,line cap=round] (226.21, 45.72) rectangle (231.74, 63.38);

\path[draw=drawColor,line width= 0.4pt,line join=round,line cap=round] (231.74, 45.72) rectangle (237.27, 56.22);

\path[draw=drawColor,line width= 0.4pt,line join=round,line cap=round] (237.27, 45.72) rectangle (242.80, 50.55);

\path[draw=drawColor,line width= 0.4pt,line join=round,line cap=round] (242.80, 45.72) rectangle (248.33, 48.61);

\path[draw=drawColor,line width= 0.4pt,line join=round,line cap=round] (248.33, 45.72) rectangle (253.86, 47.50);

\path[draw=drawColor,line width= 0.4pt,line join=round,line cap=round] (253.86, 45.72) rectangle (259.38, 46.89);

\path[draw=drawColor,line width= 0.4pt,line join=round,line cap=round] (259.38, 45.72) rectangle (264.91, 46.22);

\path[draw=drawColor,line width= 0.4pt,line join=round,line cap=round] (264.91, 45.72) rectangle (270.44, 46.00);

\path[draw=drawColor,line width= 0.4pt,line join=round,line cap=round] (270.44, 45.72) rectangle (275.97, 45.72);

\path[draw=drawColor,line width= 0.4pt,line join=round,line cap=round] (275.97, 45.72) rectangle (281.50, 45.72);

\path[draw=drawColor,line width= 0.4pt,line join=round,line cap=round] (281.50, 45.72) rectangle (287.03, 45.78);

\path[draw=drawColor,line width= 0.4pt,line join=round,line cap=round] (287.03, 45.72) rectangle (292.56, 45.78);
\end{scope}
\begin{scope}
\path[clip] (  0.00,  0.00) rectangle (469.75,216.81);
\definecolor{drawColor}{RGB}{0,0,0}

\path[draw=drawColor,line width= 0.4pt,line join=round,line cap=round] (193.04, 40.39) -- (289.79, 40.39);

\path[draw=drawColor,line width= 0.4pt,line join=round,line cap=round] (193.04, 40.39) -- (193.04, 36.43);

\path[draw=drawColor,line width= 0.4pt,line join=round,line cap=round] (206.86, 40.39) -- (206.86, 36.43);

\path[draw=drawColor,line width= 0.4pt,line join=round,line cap=round] (220.68, 40.39) -- (220.68, 36.43);

\path[draw=drawColor,line width= 0.4pt,line join=round,line cap=round] (234.50, 40.39) -- (234.50, 36.43);

\path[draw=drawColor,line width= 0.4pt,line join=round,line cap=round] (248.33, 40.39) -- (248.33, 36.43);

\path[draw=drawColor,line width= 0.4pt,line join=round,line cap=round] (262.15, 40.39) -- (262.15, 36.43);

\path[draw=drawColor,line width= 0.4pt,line join=round,line cap=round] (275.97, 40.39) -- (275.97, 36.43);

\path[draw=drawColor,line width= 0.4pt,line join=round,line cap=round] (289.79, 40.39) -- (289.79, 36.43);

\node[text=drawColor,anchor=base,inner sep=0pt, outer sep=0pt, scale=  0.66] at (193.04, 26.14) {0};

\node[text=drawColor,anchor=base,inner sep=0pt, outer sep=0pt, scale=  0.66] at (206.86, 26.14) {5};

\node[text=drawColor,anchor=base,inner sep=0pt, outer sep=0pt, scale=  0.66] at (220.68, 26.14) {10};

\node[text=drawColor,anchor=base,inner sep=0pt, outer sep=0pt, scale=  0.66] at (234.50, 26.14) {15};

\node[text=drawColor,anchor=base,inner sep=0pt, outer sep=0pt, scale=  0.66] at (248.33, 26.14) {20};

\node[text=drawColor,anchor=base,inner sep=0pt, outer sep=0pt, scale=  0.66] at (262.15, 26.14) {25};

\node[text=drawColor,anchor=base,inner sep=0pt, outer sep=0pt, scale=  0.66] at (275.97, 26.14) {30};

\node[text=drawColor,anchor=base,inner sep=0pt, outer sep=0pt, scale=  0.66] at (289.79, 26.14) {35};

\path[draw=drawColor,line width= 0.4pt,line join=round,line cap=round] (189.06, 45.72) -- (189.06,179.01);

\path[draw=drawColor,line width= 0.4pt,line join=round,line cap=round] (189.06, 45.72) -- (185.10, 45.72);

\path[draw=drawColor,line width= 0.4pt,line join=round,line cap=round] (189.06, 90.15) -- (185.10, 90.15);

\path[draw=drawColor,line width= 0.4pt,line join=round,line cap=round] (189.06,134.58) -- (185.10,134.58);

\path[draw=drawColor,line width= 0.4pt,line join=round,line cap=round] (189.06,179.01) -- (185.10,179.01);

\node[text=drawColor,rotate= 90.00,anchor=base,inner sep=0pt, outer sep=0pt, scale=  0.66] at (179.55, 45.72) {0.00};

\node[text=drawColor,rotate= 90.00,anchor=base,inner sep=0pt, outer sep=0pt, scale=  0.66] at (179.55, 90.15) {0.05};

\node[text=drawColor,rotate= 90.00,anchor=base,inner sep=0pt, outer sep=0pt, scale=  0.66] at (179.55,134.58) {0.10};

\node[text=drawColor,rotate= 90.00,anchor=base,inner sep=0pt, outer sep=0pt, scale=  0.66] at (179.55,179.01) {0.15};
\end{scope}
\begin{scope}
\path[clip] (313.17,  0.00) rectangle (469.75,216.81);
\definecolor{drawColor}{RGB}{0,0,0}

\node[text=drawColor,anchor=base,inner sep=0pt, outer sep=0pt, scale=  0.66] at (399.38, 10.30) {$\lambda_2$};
\end{scope}
\begin{scope}
\path[clip] (345.64, 40.39) rectangle (453.12,184.34);
\definecolor{drawColor}{RGB}{0,0,0}

\path[draw=drawColor,line width= 0.4pt,line join=round,line cap=round] (349.62, 45.72) rectangle (357.92, 46.50);

\path[draw=drawColor,line width= 0.4pt,line join=round,line cap=round] (357.92, 45.72) rectangle (366.21, 89.34);

\path[draw=drawColor,line width= 0.4pt,line join=round,line cap=round] (366.21, 45.72) rectangle (374.50,145.76);

\path[draw=drawColor,line width= 0.4pt,line join=round,line cap=round] (374.50, 45.72) rectangle (382.80,119.62);

\path[draw=drawColor,line width= 0.4pt,line join=round,line cap=round] (382.80, 45.72) rectangle (391.09, 86.60);

\path[draw=drawColor,line width= 0.4pt,line join=round,line cap=round] (391.09, 45.72) rectangle (399.38, 65.68);

\path[draw=drawColor,line width= 0.4pt,line join=round,line cap=round] (399.38, 45.72) rectangle (407.68, 54.57);

\path[draw=drawColor,line width= 0.4pt,line join=round,line cap=round] (407.68, 45.72) rectangle (415.97, 49.68);

\path[draw=drawColor,line width= 0.4pt,line join=round,line cap=round] (415.97, 45.72) rectangle (424.26, 48.17);

\path[draw=drawColor,line width= 0.4pt,line join=round,line cap=round] (424.26, 45.72) rectangle (432.56, 46.65);

\path[draw=drawColor,line width= 0.4pt,line join=round,line cap=round] (432.56, 45.72) rectangle (440.85, 46.02);

\path[draw=drawColor,line width= 0.4pt,line join=round,line cap=round] (440.85, 45.72) rectangle (449.14, 45.87);

\path[draw=drawColor,line width= 0.4pt,line join=round,line cap=round] (449.14, 45.72) rectangle (457.44, 45.91);

\path[draw=drawColor,line width= 0.4pt,line join=round,line cap=round] (457.44, 45.72) rectangle (465.73, 45.87);

\path[draw=drawColor,line width= 0.4pt,line join=round,line cap=round] (465.73, 45.72) rectangle (474.02, 45.80);
\end{scope}
\begin{scope}
\path[clip] (  0.00,  0.00) rectangle (469.75,216.81);
\definecolor{drawColor}{RGB}{0,0,0}

\path[draw=drawColor,line width= 0.4pt,line join=round,line cap=round] (349.62, 40.39) -- (449.14, 40.39);

\path[draw=drawColor,line width= 0.4pt,line join=round,line cap=round] (349.62, 40.39) -- (349.62, 36.43);

\path[draw=drawColor,line width= 0.4pt,line join=round,line cap=round] (366.21, 40.39) -- (366.21, 36.43);

\path[draw=drawColor,line width= 0.4pt,line join=round,line cap=round] (382.80, 40.39) -- (382.80, 36.43);

\path[draw=drawColor,line width= 0.4pt,line join=round,line cap=round] (399.38, 40.39) -- (399.38, 36.43);

\path[draw=drawColor,line width= 0.4pt,line join=round,line cap=round] (415.97, 40.39) -- (415.97, 36.43);

\path[draw=drawColor,line width= 0.4pt,line join=round,line cap=round] (432.56, 40.39) -- (432.56, 36.43);

\path[draw=drawColor,line width= 0.4pt,line join=round,line cap=round] (449.14, 40.39) -- (449.14, 36.43);

\node[text=drawColor,anchor=base,inner sep=0pt, outer sep=0pt, scale=  0.66] at (349.62, 26.14) {0};

\node[text=drawColor,anchor=base,inner sep=0pt, outer sep=0pt, scale=  0.66] at (366.21, 26.14) {10};

\node[text=drawColor,anchor=base,inner sep=0pt, outer sep=0pt, scale=  0.66] at (382.80, 26.14) {20};

\node[text=drawColor,anchor=base,inner sep=0pt, outer sep=0pt, scale=  0.66] at (399.38, 26.14) {30};

\node[text=drawColor,anchor=base,inner sep=0pt, outer sep=0pt, scale=  0.66] at (415.97, 26.14) {40};

\node[text=drawColor,anchor=base,inner sep=0pt, outer sep=0pt, scale=  0.66] at (432.56, 26.14) {50};

\node[text=drawColor,anchor=base,inner sep=0pt, outer sep=0pt, scale=  0.66] at (449.14, 26.14) {60};

\path[draw=drawColor,line width= 0.4pt,line join=round,line cap=round] (345.64, 45.72) -- (345.64,179.01);

\path[draw=drawColor,line width= 0.4pt,line join=round,line cap=round] (345.64, 45.72) -- (341.68, 45.72);

\path[draw=drawColor,line width= 0.4pt,line join=round,line cap=round] (345.64, 90.15) -- (341.68, 90.15);

\path[draw=drawColor,line width= 0.4pt,line join=round,line cap=round] (345.64,134.58) -- (341.68,134.58);

\path[draw=drawColor,line width= 0.4pt,line join=round,line cap=round] (345.64,179.01) -- (341.68,179.01);

\node[text=drawColor,rotate= 90.00,anchor=base,inner sep=0pt, outer sep=0pt, scale=  0.66] at (336.14, 45.72) {0.00};

\node[text=drawColor,rotate= 90.00,anchor=base,inner sep=0pt, outer sep=0pt, scale=  0.66] at (336.14, 90.15) {0.03};

\node[text=drawColor,rotate= 90.00,anchor=base,inner sep=0pt, outer sep=0pt, scale=  0.66] at (336.14,134.58) {0.06};

\node[text=drawColor,rotate= 90.00,anchor=base,inner sep=0pt, outer sep=0pt, scale=  0.66] at (336.14,179.01) {0.09};
\end{scope}
\end{tikzpicture}
\end{center}
\caption{Histogram of the posterior samples}
\end{figure}

\vspace*{5mm}

\begin{table}[!ht]\small
\caption{Confidence and credible intervals for the parameters}   \label{data-ci}
\begin{center}
\begin{tabular}{*{6}{c}}
\toprule
Parameter & Nominal CL & BC-bootstrap & P-bootstrap & Sym. CRI & HPD CRI \\
\midrule
\multirow{2}{*}{$\alpha$}    & 90\% & (2.204, 3.264)   & (2.343, 3.439) & (2.250, 3.342) & (2.247, 3.332)\\
		                     & 95\% & (2.102, 3.366)   & (2.264, 3.529) & (2.161, 3.462) & (2.153, 3.446)\\
\midrule
\multirow{2}{*}{$\lambda_1$} & 90\% & (0.000, 11.694) & (3.601, 15.324) & (2.892, 13.970) & (2.129, 11.924)\\
                    		 & 95\% & (0.000, 12.856) & (3.018, 17.795) & (2.534, 16.167) & (1.657, 14.060)\\
\midrule
\multirow{2}{*}{$\lambda_2$} & 90\% & (0.000, 27.567) & (8.925, 35.127) & (7.725, 30.963) & (5.506, 26.916)\\
                 		     & 95\% & (0.000, 30.340) & (7.755, 41.151) & (6.838, 35.697) & (5.397, 31.623)\\
\bottomrule
\end{tabular}
\end{center}
\end{table}


\section{\sc Shape Parameters are Different} 

\subsection{\sc Classical Inference}

So far we have provided the analysis based on the assumption that the shape parameters of the lifetime distributions of the competing 
causes are same.  In this section we relax that assumption.  It is assumed that $T_{1i} \sim$ Weibull($\alpha_1, \lambda_1$), and 
$T_{2i} \sim$ Weibull($\alpha_2, \lambda_2$), and they are independent.  We are using the same notations as before.  The likelihood 
function in this case can be written as
$$
L_2(\alpha_1, \lambda_1, \alpha_2, \lambda_2) = \alpha_1^{m_1} \alpha_2^{m_2} \lambda_1^{m_1} \lambda_2^{m_2}  
e^{-\lambda_1 \left [ \sum_{i=1}^n (t_i^{\alpha_1} - (1-\nu_i)\tau_{iL}^{\alpha_1}) \right ]} 
 e^{-\lambda_2 \left [ \sum_{i=1}^n (t_i^{\alpha_2} -(1-\nu_i) \tau_{iL}^{\alpha_2}) \right ]} \prod_{i \in I_1} t_i^{\alpha_1-1}
\prod_{i \in I_2} t_i^{\alpha_2-1}.
$$
Hence, the log-likelihood function can be written as
\begin{eqnarray}
\log L_2(\alpha_1, \alpha_2, \lambda_1, \lambda_2) & = & m_1 \log \alpha_1 + m_1 \log \lambda_1 + (\alpha_1-1) w_1^* - \lambda_1 w_2(\alpha_1) 
\nonumber \\
 &  & + m_2 \log \alpha_2 + m_2 \log \lambda_2 + (\alpha_2-1) w_2^* - \lambda_2 w_2(\alpha_2),    \label{lik-diff}
\end{eqnarray} 
where
$$
w_1^* = \sum_{i \in I_1} t_i, \ \ \ w_2^* = \sum_{i \in I_2} t_i, 
$$
and $w_2(\alpha)$ is same as defined in (\ref{w1w2}).
Following the same approach as in Section 3, it immediately follows that for fixed $\alpha_1$ and $\alpha_2$, the MLEs of $\lambda_1$ 
and $\lambda_2$ can be obtained as:
\[
\widehat{\lambda}_1(\alpha_1) = \frac{m_1}{w_2(\alpha_1)}, \quad \widehat{\lambda}_2(\alpha_2) = \frac{m_2}{w_2(\alpha_2)},
\]
and the MLEs of $\alpha_1$ and $\alpha_2$ can be obtained by maximizing 
\begin{align*}
p_1(\alpha_1) = m_1 \log \alpha_1 - m_1 \log w_2(\alpha_1) + \alpha_1 w_1^* \\
\shortintertext{and}
p_2(\alpha_2) = m_2 \log \alpha_2 - m_2 \log w_2(\alpha_2) + \alpha_2 w_2^*,
\end{align*}
with respect to $\alpha_1$ and $\alpha_2$, respectively. Following Lemma 2, it can be shown that
$p_1(\alpha_1)$ and $p_2(\alpha_2)$ are unimodal functions if $d(\alpha)\geq0$ for all $\alpha>0$. Hence, the
corresponding maximization can be performed quite conveniently. One can employ a numerical technique, such as the
Newton-Raphson to obtain the numerical estimates of $\alpha_1$ and $\alpha_2$ from (\ref{lik-diff}). Then, the obtained
estimates $\widehat{\alpha}_1$ and $\widehat{\alpha}_2$ can be plugged in to obtain the estimates
$\widehat{\lambda}_1(\widehat{\alpha}_1)$ and $\widehat{\lambda}_2(\widehat{\alpha}_2)$.  Parametric bootstrap
confidence intervals can be obtained similarly as in Section 3.

\subsection{\sc Bayesian Inference}

We make the following prior assumptions on the unknown parameters:
$$
\pi(\alpha_1) \sim \hbox{GA}(a_1,b_1), \ \ \ \pi(\lambda_1) \sim \hbox{GA}(c_1,d_1), \ \ \ \ \pi(\alpha_2) \sim \hbox{GA}(a_2,b_2), 
\ \ \ \ \pi(\lambda_2) \sim \hbox{GA}(c_2,d_2).
$$
It follows that the posterior distributions of $(\alpha_1,\lambda_1)$ is independent of the posterior distribution of 
$(\alpha_2,\lambda_2)$, i.e.
$$
\pi(\alpha_1, \lambda_1, \alpha_2, \lambda_2|data) = \pi(\alpha_1, \lambda_1|data) \times \pi(\alpha_2, \lambda_2|data)
$$
It can be shown after some calculations that 
$\displaystyle \pi(\lambda_1|\alpha_1, data)  \sim \hbox{GA}(m_1+c_1, w_2(\alpha_1)+d_1)$,
$$
\pi(\alpha_1|data) \propto \alpha_1^{m_1+a_1-1} e^{-b_1 \alpha_1} \prod_{i \in I_1} t_i^{\alpha_1-1} \times
\frac{1}{(w_2(\alpha_1)+d_1)^{m_1+c_1}}; 
\ \ \ \hbox{for} \ \ \ \alpha_1 > 0.
$$
and
$\displaystyle \pi(\lambda_2|\alpha_2, data) \sim \hbox{GA}(m_2+c_2, w_2(\alpha_2)+d_2)$,
$$
\pi(\alpha_2|data) \propto \alpha_2^{m_2+a_2-1} e^{-b_2 \alpha_2} \prod_{i \in I_2} t_i^{\alpha_2-1} \times
\frac{1}{(w_2(\alpha_2)+d_2)^{m_2+c_2}}; 
\ \ \ \hbox{for} \ \ \ \alpha_2 > 0.
$$
Using Lemma 3, it immediately follows that for $i=1,\,2$; $\displaystyle \pi(\alpha_i| data)$ is log-concave function if
$w_2^{\prime\prime}(\alpha_i)(w_2(\alpha_i)+d_i)-\left\{ w_2^\prime(\alpha_i) \right\}^2\geq 0$. In this case, the
generations from the posterior distribution can be performed quite conveniently, and once we have the generated samples,
the Bayes estimate of any function of the unknown parameters and the associated credible intervals can be easily
constructed as before.

\subsection{\sc Transformer Data Revisited}

We re-analyze the same transformer data set under the new set of assumptions.  The MLEs of the unknown parameters are 
as follows: 
$$
\widehat{\alpha}_1 = 2.817 \quad \widehat{\lambda}_1 = 6.933, \ \ \ 
\widehat{\alpha}_2 = 2.786 \quad \widehat{\lambda}_2 = 15.768.
$$
The Bayes estimates based on the non-informative priors of the unknown parameters turn out to be:
$$
\widehat{\alpha}_1 = 2.776 \quad \widehat{\lambda}_1 = 8.532, \ \ \ 
\widehat{\alpha}_2 = 2.767 \quad \widehat{\lambda}_2 = 17.083.
$$
Different bootstrap confidence intervals and credible intervals for the parameters are given in the Table \ref{sh-diff} below.
\begin{table}[!ht]\small
	\caption{Confidence and credible intervals for the parameters   \label{sh-diff}}   
	\begin{center}
		\begin{tabular}{*{6}{c}}
			\toprule
			Parameter & Nominal CL & BC-bootstrap & P-bootstrap & Sym.~CRI & HPD CRI\\
			\midrule
			\multirow{2}{*}{$\alpha_1$}    & 90\% & (1.696, 3.726)    & (1.927, 3.963)  & (1.842, 3.836)  & (1.792, 3.762) \\
			                               & 95\% & (1.501, 3.921)    & (1.775, 4.287)  & (1.683, 4.049)  & (1.648, 4.005) \\
			\midrule                                                                                          
			\multirow{2}{*}{$\alpha_2$}    & 90\% & (1.999, 3.347)    & (2.278, 3.557)  & (2.148, 3.440)  & (2.126, 3.408) \\
			                               & 95\% & (1.870, 3.476)    & (2.195, 3.830)  & (2.033, 3.566)  & (1.990, 3.519) \\
			\midrule                                                                                          
			\multirow{2}{*}{$\lambda_1$}   & 90\% & (-10.933, 18.678) & (2.173, 24.150) & (1.875, 22.696) & (0.923, 17.126)\\
			                               & 95\% & (-13.770, 21.514) & (1.766, 34.984) & (1.544, 28.991) & (0.577, 22.787)\\
			\midrule                                                                                          
			\multirow{2}{*}{$\lambda_2$}   & 90\% & (-9.715, 32.246)  & (8.073, 39.503) & (6.884, 34.509) & (4.935, 29.175)\\
			                               & 95\% & (-13.735, 36.265) & (6.608, 48.713) & (5.919, 40.036) & (4.077, 34.845)\\
			\bottomrule
		\end{tabular}
	\end{center}
\end{table}

We perform a hypothesis test, to check whether it is possible to take the shape parameters to be equal for this data.  We would 
like to test
\[
H_0: \alpha_1 = \alpha_2, \quad \textrm{against} \quad H_1: \alpha_1 \ne \alpha_2,
\]     
based on the likelihood ratio test procedure. We observe that likelihood ratio test statistic is 
\[
\chi^2 = -2(\sup_{H_0} \log L_1(\alpha, \lambda_1, \lambda_2) - \sup_{H_0 \cup H_1} \log L_2(\alpha_1, \alpha_2, \lambda_1, \lambda_2)) = 0.0018
\] 
Comparing the test statistic with $\chi^2_1(0.95) = 3.841$, we conclude that $H_0$ cannot be rejected at $5\%$ level.

\section{\sc Conclusions}

In this paper we consider the classical and Bayesian inference of the Weibull parameters for the left truncated and right censored
competing risks data.  We provide the sufficient condition of the existence and uniqueness of the maximum likelihood estimators of
the unknown parameters.  It is difficult to obtain the exact confidence intervals of the unknown parameters and we propose to use
the bootstrap method to construct the approximate confidence intervals.  We have further considered the Bayesian inference of the
unknown parameters under a very flexible priors on the unknown parameters.  We propose to use the importance sampling procedure to
compute the Bayes estimates and the associated credible intervals.  Extensive simulation experiments have been performed to
compare the performances of the estimation procedures for the unknown parameters, and it is observed that the performance of the Bayes estimates with
non-informative priors are slightly better than the maximum likelihood estimators.  Finally we extend the results when the two
shape parameters are different.  We provide both the classical and Bayesian inference of the unknown parameters.  We re-analyze
the same transformer data set and it is observed that for this data set the assumption of equal shape parameters makes sense.  It
will be of interest to consider the case when there are some covariates also associated with each item.  Moreover, prior
elicitation is an important problem which has not addressed in this paper.  More work is needed in these directions.

\section*{\sc Acknowledgements:} The authors would like to thank the associate editor and two referees for their constructive comments
which have helped to improve the manuscript significantly.

\section*{References}
\begin{description}
\item Balakrishnan N., Kundu D., Ng H. K. T., and Kannan N. (2007). Point and interval estimation for a simple step-stress model
with Type-II censoring. {\it Journal of Quality Technology}, {\bf 39}, 35 - 47.

\item Balakrishnan, N.; Mitra, D. (2011). Likelihood inference for log-normal data with left truncation and right censoring with 
an illustration, {\it Journal of Statistical Planning and Inference}, \textbf{141}, 3536 - 3553.

\item Balakrishnan, N. and Mitra, D. (2012). Left truncated and right censored Weibull data and likelihood inference with an 
illustration. {\it Computational Statistics and Data Analysis}, \textbf{56}, 4011 - 4025. 

\item Balakrishnan N. and Mitra D. (2014). EM-based likelihood inference for some lifetime distributions based on left truncated
and right censored data and associated model discrimination. {\it South African Statistical Journal}, {\bf 48}, 125 - 171.

\item Cox D. R. (1959). The analysis of exponentially distributed lifetimes with two types of failures. {\it Journal of the Royal
Statistical Society Series B}, {\bf 21}, 411 - 421.

\item Crowder M. (2001). {\it Classical Competing Risks Model}. Chapman \& Hall, New York.

\item David, H.A. and Moeschberger, M.L. (1978). {\it The Theory of Competing Risks}, Griffin, London.

\item Hong Y., Meeker W. Q., and McCalley J. D. (2009). Prediction of remaining life of power transformers based on left truncated
and right censored lifetime data. {\it The Annals of Applied Statistics}, {\bf 3}, 857-879.

\item Devroye L. (1984). A simple algorithm for generating random variates with a log-concave density. {\it Computing}, {\bf 33},
247 - 257.

\item Kinderman, A.J. and Monahan, J.F. (1977). Computer generation of random variables using the ratio of random deviates,
\textit{ACM Transactions on Mathematical Software}, \textbf{3}, 257 - 260.

\item Kundu D. (2004). Parameter estimation for partially complete time and type of failure data. {\it Biometrical Journal}, {\bf
46}, 165 - 179.

\item Kundu D. (2008). Bayes inference and life testing plan for the Weibull distribution in presence of progressive censoring.
{\it Technometrics}, {\bf 50}, 144 - 154.

\item Kundu, D. and Mitra, D. (2016). Bayesian inference of Weibull distribution based on left truncated and right censored data, 
{\it Computational Statistics and Data Analysis}, \textbf{99}, 38-50. 

\item Kundu D. and Pradhan B. (2011). Bayesian analysis of progressively censored competing risks data. {\it Sankhya Series B},
{\bf 73}, 276 - 296.


\item Pena E. A. and Gupta A. K. (1990). Bayes estimation for the Marshall-Olkin exponential distribution. {\it Journal of the
Royal Statistical Society Series B}, {\bf 52}, 379 - 389.

\item Prentice R. L., Kalbfleish J. D., Peterson Jr. A. V., Flurnoy N., Farewell V. T., Breslow N. E. (1978). The analysis of
failure times in presence of competing risks. {\it Biometrics}, {\bf 34}, 541 - 554.


\end{description}

\begin{appendices}
\section{Transformer Data}
\begin{table}\footnotesize
\caption{Year of installation, year of exit of the transformers along with the truncation, censoring indicator and cause of
failures} 
\begin{center}
\begin{tabular}{|r|c|c|c|c||r|c|c|c|c||r|c|c|c|c|}
\toprule
S.N. & Year of & Year of & $\nu$ & $\delta$ & S.N. & Year of & Year of & $\nu$ & $\delta$  & S.N. & Year of & Year of & $\nu$ & $\delta$    \\   
     & Inst.  & Exit &  &  &  & Inst. & Exit &  &  &  & Inst.  & Exit &  &  \\  \midrule \midrule
      1  &       1961&   1996  &     0&    2  &     11  &       1963&   2008  &     0&    0 &     21  &       1960&   1988  &     0&    1  \\
      2  &       1964&   1985  &     0&    1   &    12  &       1963&   2000  &     0&    1 &     22  &       1961&   1993  &     0&    2  \\
      3  &       1962&   2007  &     0&    2  &    13  &       1960&   1981  &     0&    2 &     23  &       1961&   1990  &     0&    2  \\
      4  &       1962&   1986  &     0&    2   &    14  &       1963&   1984  &     0&    2&     24  &       1960&   1986  &     0&    1  \\
      5  &       1961&   1992  &     0&    2  &    15  &       1963&   1993  &     0&    2 &     25  &       1962&   2008  &     0&    0  \\
      6  &       1962&   1987  &     0&    1  &    16  &       1964&   1992  &     0&    2 &     26  &       1964&   1982  &     0&    2  \\ 
      7  &       1964&   1993  &     0&    2  &    17  &       1961&   1981  &     0&    2 &     27  &       1963&   1984  &     0&    1  \\ 
      8  &       1960&   1984  &     0&    2  &    18  &       1960&   1995  &     0&    1 &     28  &       1960&   1987  &     0&    2  \\ 
      9  &       1963&   1997  &     0&    2 &    19  &       1961&   2008  &     0&    0  &     29  &       1962&   1996  &     0&    2  \\  
     10  &       1962&   1995  &     0&    2  &    20  &       1960&   2002  &     0&    1 &      30  &       1963&   1994  &     0&    1  \\  
\midrule \midrule
     31  &       1987&   2008  &     1&    0  &   41  &       1980&   2008  &     1&    0 &     51  &       1984&   2001  &     1&    2  \\
     32  &       1980&   2008  &     1&    0  &   42  &       1982&   2008  &     1&    0 &     52  &       1983&   2008  &     1&    0  \\
     33  &       1988&   2008  &     1&    0  &   43  &       1986&   2008  &     1&    0 &     53  &       1988&   2008  &     1&    0  \\ 
     34  &       1985&   2008  &     1&    0  &     44  &       1984&   2008  &     1&  0 &     54  &       1988&   2008  &     1&    0  \\ 
     35  &       1989&   2008  &     1&    0  &     45  &       1986&   1995  &     1&  2 &     55  &       1985&   2008  &     1&    0  \\ 
     36  &       1981&   2008  &     1&    0  &     46  &       1986&   2008  &     1&  0 &     56  &       1986&   2008  &     1&    0  \\ 
     37  &       1985&   2008  &     1&    0 &     47  &       1987&   2008  &     1&   0 &     57  &       1988&   2008  &     1&    0  \\ 
     38  &       1986&   2004  &     1&    2 &     48  &       1986&   2008  &     1&    0 &     58  &       1982&   2008  &     1&    0  \\ 
     39  &       1980&   1987  &     1&    2 &     49  &       1986&   2008  &     1&    0 &     59  &       1985&   2008  &     1&    0  \\
     40  &       1986&   2005  &     1&    1 &     50  &       1984&   2008  &     1&    0 &     60  &       1988&   2008  &     1&    0  \\ 
\midrule \midrule
    61  &       1982&   2004  &     1&    2  &     71  &       1989&   2008  &     1&    0   &    81  &       1981&   2006  &     1&    2  \\
     62  &       1980&   2008  &     1&    0  &     72  &       1989&   2008  &     1&    0  &    82  &       1988&   1996  &     1&    1  \\  
     63  &       1980&   2002  &     1&    2  &     73  &       1986&   2008  &     1&    0   &    83  &       1985&   2002  &     1&    2  \\
     64  &       1984&   2008  &     1&    0  &     74  &       1982&   1999  &     1&    2  &    84  &       1984&   2008  &     1&    0  \\
     65  &       1981&   1999  &     1&    1  &     75  &       1985&   2008  &     1&    0  &    85  &       1980&   2008  &     1&    0  \\
     66  &       1986&   2007  &     1&    2  &     76  &       1986&   2008  &     1&    0  &    86  &       1982&   2008  &     1&    0  \\
     67  &       1987&   2008  &     1&    0  &     77  &       1982&   2008  &     1&    0  &    87  &       1981&   1995  &     1&    2  \\
     68  &       1983&   2008  &     1&    0  &     78  &       1988&   2004  &     1&    1  &    88  &       1986&   1997  &     1&    2  \\
     69  &       1983&   2006  &     1&    2  &     79  &       1980&   2008  &     1&    0  &    89  &       1986&   2008  &     1&    0  \\
     70  &       1983&   1993  &     1&    1  &     80  &       1982&   2002  &     1&    2  &    90  &       1986&   2008  &     1&    0  \\ 
\midrule  \midrule
     91  &       1982&   2008  &     1&    0  &     96  &       1986&   2008  &     1&    0  &  &  &  &  & \\  
     92  &       1989&   2008  &     1&    0  &     97  &       1982&   1996  &     1&    2   &  &  &  &  & \\
     93  &       1984&   2008  &     1&    0  &     98  &       1982&   2008  &     1&    0  &  &  &  &  &  \\
     94  &       1980&   2008  &     1&    0  &     99  &       1982&   2008  &     1&    0  &  &  &  &  &  \\
     95  &       1988&   2008  &     1&    0  &    100  &       1989&   2008  &     1&    0   &  &  &  &  &  \\
\bottomrule
\end{tabular}
\end{center}
\label{sim-data}
\end{table}

\end{appendices}

\end{document}